\documentclass[showpacs,preprintnumbers,
amsmath,amssymb,groupedaddress,superscriptaddress]{revtex4}
\usepackage{graphicx}
\usepackage{dcolumn}
\usepackage{bm}

\begin{document}
\draft
\title{Nuclear isospin mixing and elastic parity-violating electron scattering}

\author{O. Moreno}
\address{Instituto de Estructura de la Materia,
CSIC, Serrano 123, E-28006 Madrid, Spain}
\author{P. Sarriguren}
\address{Instituto de Estructura de la Materia,
CSIC, Serrano 123, E-28006 Madrid, Spain}
\author{E. Moya de Guerra}
\address{Departamento de F\'isica At\'omica, Molecular y Nuclear,
Universidad Complutense de Madrid, \\ E-28040 Madrid, Spain}
\author{J.M. Udias}
\address{Departamento de F\'isica At\'omica, Molecular y Nuclear,
Universidad Complutense de Madrid, \\ E-28040 Madrid, Spain}
\author{T.W. Donnelly}
\address{Center for Theoretical Physics, Laboratory for Nuclear Science
and Department of Physics, \\
Massachusetts Institute of Technology, Cambridge, MA 02139, USA}
\author{I. Sick}
\address{Departement f\"ur Physik, Universit\"at Basel, CH-4056 Basel, Switzerland}

%\date{\today}

\begin{abstract}

  The influence of nuclear isospin mixing on parity-violating elastic
  electron scattering is studied for the even-even, $N=Z$ nuclei
  $^{12}$C, $^{24}$Mg, $^{28}$Si, and $^{32}$S. Their ground-state
  wave functions have been obtained using a self-consistent
  axially-symmetric mean-field approximation with density-dependent
  effective two-body Skyrme interactions. Some differences from
  previous shell-model calculations appear for the isovector Coulomb
  form factors which play a role in determining the parity-violating
  asymmetry.  To gain an understanding of how these differences arise,
  the results have been expanded in a spherical harmonic oscillator
  basis.  Results are obtained not only within the plane-wave Born
  approximation, but also using the distorted-wave Born approximation
  for comparison with potential future experimental studies of
  parity-violating electron scattering. To this end, for each nucleus
  the focus is placed on kinematic ranges where the signal
  (isospin-mixing effects on the parity-violating asymmetry) and the
  experimental figure-of-merit are maximized.  Strangeness
  contributions to the asymmetry are also briefly discussed, since
  they and the isospin mixing contributions may play comparable roles
  for the nuclei being studied at the low momentum transfers of
  interest in the present work.

\end{abstract}


\maketitle

\section{Introduction}

One motivation for studying parity-violating (PV) electron scattering
from nuclei is to use it as a tool to extract information on the weak
neutral current (WNC) and thereby to test the validity of the Standard
Model in the low-energy regime~\cite{Fei75,Wal77} (see also
\cite{Don79,Mus94}). This possibility lies in the fact that the PV
asymmetry acquires a very simple, model-independent expression in
terms of basic coupling constants, with nuclear structure effects
cancelling out, if certain conditions are met. The assumptions
required to arrive to such a simple expression are: (1) that the focus
is placed on elastic scattering from spin-zero nuclear targets, for
then only Coulomb-type monopole form factors enter; (2) that
strangeness content in the WNC can be neglected, for then only
isoscalar and isovector matrix elements occur, with no third type of
contribution; and (3) that the nuclear ground states have isospin
zero, permitting only a single Coulomb monopole matrix element to
occur, namely, the isoscalar one. In fact, while restriction (1) can
be met with a wide range of even-even nuclei, restrictions (2) and (3)
are not completely attainable. Strangeness content in the WNC has been
extensively studied in previous work (see, for instance, the review
article~\cite{Mus94}) and is not the primary focus of the present
study. Instead, the present work is aimed at the last issue of
potential isospin mixing in nuclear ground states and how this affects
the PV asymmetry, albeit taking into account present uncertainties in
the strangeness content of the nucleon.

Isospin mixing constituted part of the goal of a previous study by
Donnelly, Dubach and Sick (DDS)~\cite{Don89} where the effect of
isospin mixing in nuclear ground states on PV electron scattering was
first studied. The approach taken there was to use a simple two-level
model in which the $T=0$ ground state and an excited $T=1$ state, both
with angular momentum/parity 0$^+$, were admixed with a mixing
parameter $\chi$. The deviation in the PV asymmetry due to isospin
mixing was found to be proportional to this mixing parameter $\chi$
and to the ratio between the inelastic isovector and the elastic
isoscalar Coulomb monopole form factors. In DDS a shell-model approach
with only one active shell in a spherical harmonic oscillator basis
was used to obtain the spin-isospin reduced Coulomb matrix elements
(isoscalar and isovector) for $^{12}$C and $^{28}$Si. Coulomb
distortion effects on the electron waves were also neglected in that
earlier work.

Isospin dependence in PV electron scattering is crucial in attempting
to determine the precision up to which the Standard Model constants
can be deduced and to what extent strangeness effects in the WNC can
be studied. At the same time, this dependence can be exploited to
provide information about the spatial distribution of neutrons in the
nuclear ground state. Indeed, this idea was proposed as part of the
original study by DDS, namely that a measurement of the
electron-scattering PV-asymmetry can provide a direct measurement of
the Fourier Transform of the neutron density. In a subsequent
study~\cite{Hor01} the idea in DDS was extended to allow for Coulomb
distortion of the electron wave function, something also done in the
present work. Neutron densities are less well known than charge
densities (studied using parity-conserving electron scattering) as
most of this information comes from hadronic probes where the reaction
mechanism involved is more difficult to interpret than is the case for
semi-leptonic electroweak processes. Thus, an electroweak measurement
of the neutron density can serve to calibrate the measurements made
with hadronic probes. In fact, the idea proposed by DDS is now being
realized in the Parity Radius Experiment (PREX) at Jefferson
Laboratory which has the goal of measuring the neutron radius of
$^{208}$Pb using PV elastic electron scattering. Such a measurement
has implications for astrophysics (including the structure of neutron
stars), for atomic parity non-conservation studies, for the structure
of neutron-rich nuclei, and for determinations of neutron skin
thicknesses of nuclei.

In this work we extend the previous study undertaken by DDS to examine
the effects on the PV asymmetry induced by isospin admixtures in the
nuclear ground states of N=Z nuclei occurring when Coulomb
interactions between nucleons are taken into account and pairing
correlations are included. The nuclear structure is described within a
self-consistent axially-symmetric Hartree-Fock formalism with Skyrme
forces; in fact, three different forces are used to explore the
sensitivity of the PV asymmetry deviation to the nuclear dynamics.
Several additional extensions beyond the work of DDS are made, namely,
inclusion of the spin-orbit correction to the Coulomb monopole
operators, use of modern electroweak form factors for protons and
neutrons, inclusion of strangeness contributions in the nucleon form
factors, and, as noted above, incorporation of the effects caused by
the Coulomb distortion of the electron wave functions. The present
study includes four N=Z isotopes, namely, $^{12}$C, $^{24}$Mg,
$^{28}$Si, and $^{32}$S.

With respect to nuclear structure issues, the main difference between
the treatment in DDS and here is that in DDS Coulomb effects in the
nuclear ground state are considered perturbatively to find the
admixture of giant isovector monopole strength ($J=0^+$, $T=1$), as
explained at length in \cite{Aue83}. Here, within the self-consistent
mean-field approach with two-body density-dependent effective
interactions, the collective isospin mixing effect of the Coulomb
force is included non-perturbatively in the isospin-non-conserving
Hartree-Fock (HF) mean field, along with other collective effects such
as pairing and deformation. As a result, the HF+BCS ground state is
made up of quasiparticles with rather complex admixtures of harmonic
oscillator wave functions in many different major shells.

In discussing the results one sees that there are limited kinematical
regions where the PV asymmetry is measurable at the levels required,
{\it i.e.,} as characterized by the figure-of-merit (FOM) which
typically peaks at momentum transfers around 0.5 fm$^{-1}$. For the
region extending up to a little beyond 1 fm$^{-1}$ the effects of
isospin mixing on the PV asymmetry will be seen later to be
characteristically at the level of a few percent for light nuclei, up
to more than 10\% for the heavier nuclei considered. These are
certainly within the scope of existing and future measurements of PV
electron scattering.  For example, the HAPPEX-He measurements at JLab
\cite{Acha07} have already been performed at low $q$, {\it i.e.,} for
kinematics similar to those of relevance in the present work. These
yielded PV asymmetries at the 4\% level (which are statistically, not
systematically limited at present), while the future PREX measurements
on Pb \cite{PREX} aim for 3\% precision in the PV asymmetry. In fact,
in recent work (see \cite{Lhu08} and references therein) it has been
noted that interpretations of results from experiments like HAPPEX and
G0 are reaching the point where they are limited by effects from
isospin mixing. For heavier nuclei, as discussed in the present work,
the effects from isospin mixing will be seen to be even larger than
for the nucleon or few-body nuclei.
  
At higher momentum transfers the effects of isospin mixing are still
larger; however, the FOM is smaller in this region and, as a
consequence, our attention in the present work is focused only on the
lower momentum transfer region.  As far as other contributions are
concerned, after discussing our basic results, and thereby setting the
scale of the isospin mixing, we return briefly to show the size of
potential strangeness effects and of the expected influence of
meson-exchange currents at these lower momentum transfers. Other
effects such as parity admixtures in the nuclear ground state,
dispersion corrections, or radiative corrections are expected to
provide negligible modifications to the asymmetry \cite{Hor01} and are
not considered in this work.

The outline of the paper is the following. In Sect. II we start by
introducing the formalism necessary to describe PV in polarized
elastic electron scattering from spin-zero N=Z nuclei. The correction
to the asymmetry due to isospin mixing is isolated and analyzed in
terms of the various ingredients involved, such as the roles played by
the nucleon form factors and by potential strangeness
contributions. There we also discuss the effects expected from
electron distortions. In Sect. III we present our results on the PV
asymmetry for the four nuclei under study. First we explore the
influence on the results of the various ingredients in the formalism
(different pairing gaps, different Skyrme interactions, the spin-orbit
correction) and go on to compare the present work with past studies of
isospin-mixing effects in PV electron scattering.  Following this, the
principal results of the study are presented for the four chosen
nuclei and finally, in Sect. IV, we summarize the main conclusions of
our work.

\section{Formalism}

\subsection{Parity-violation asymmetry}
Polarized single-arm electron scattering from unpolarized nuclei can
be used to study parity violation, since both electromagnetic (EM) and
weak interactions contribute to the process via $\gamma$ and $Z^0$
exchange, respectively.  The PV asymmetry is given by~\cite{Mus94}
\begin{equation}\label{asymmetry_sigmas}
{\mathcal A}=\frac{d\sigma^+ - d\sigma^-}{d\sigma^+ + d\sigma^-}\, ,
\end{equation}
where $d\sigma^+ (d\sigma^-)$ is the cross section for electrons
longitudinally polarized parallel (antiparallel) to their
momentum. Keeping only the square of the photon-exchange amplitude for
the spin-averaged EM cross section and using the interference between
the $\gamma$ and $Z^{0}$ amplitudes in the cross section difference,
in Plane Wave Born Approximation (PWBA) the asymmetry $\mathcal A$ in
the Standard Model can be written as
\begin{equation}
{\mathcal A}= \frac{G}{2\pi\alpha\sqrt{2}}|Q^2|\frac{W^{PV}}{F^2}\, ,
\end{equation}
where $G$ and $\alpha$ are the Fermi and fine-structure coupling
constants, respectively, and $Q$ is the four-momentum transfer in the
scattering process. $W^{PV}$ is the PV response and $F$ is the EM form
factor, both containing the dependence of the asymmetry on the nuclear
structure.  In the case of elastic electron scattering between $J^\pi
=0^+$ states, only the Coulomb-type monopole operators can induce the
transition,
\begin{equation}
F^2(q)=v_L F_{C0}^2 (q), \quad W^{PV}=v_L a_A F_{C0}(q) \tilde{F}_{C0} (q)\, ,
\end{equation}
where $a_A=-1$, $v_L$ is a kinematical factor that cancels out in the
ratio, and $F_{C0} (\tilde{F}_{C0})$ is the EM(WNC) monopole Coulomb
form factor.  Then one has
\begin{equation}\label{hadronicratio}
\frac{W^{PV}}{F^2} = a_A\frac{ \tilde{F}_{C0} (q)}{ F_{C0}(q)}\, .
\end{equation}
If we now consider only $N=Z$ nuclei and assume that they are isospin
eigenstates with isospin zero in their ground states, then only
isoscalar matrix elements contribute and the weak and EM form factors
become proportional:
\begin{equation}
\tilde{F}_{C0} (q) = \beta ^{(0)}_V F_{C0}(q)\, .
\end{equation}
Accordingly, the PV asymmetry does not depend on the form factors:
\begin{equation}
{\mathcal A} = {\mathcal A}^0 \equiv \left[ \frac{G\left| Q^2 \right|
}{2\pi \alpha \sqrt{2}} \right] a_A \beta ^{(0)}_V \cong 3.22 \times
10^{-6} |Q^2|\; \text{(in fm}^{-2}\text{)} \, ,
\label{referencevalue}
\end{equation}
where, within the Standard Model, $a_A \beta_V^{(0)}=2\sin^2\theta_W$,
$\theta_W$ being the weak mixing angle.

The actual PV asymmetry deviates from this constant value by a
correction $\Gamma$, where
\begin{eqnarray} \label{asymmetryratio}
{\mathcal A}={\mathcal A}^0(1+\Gamma)
\end{eqnarray}
or equivalently (for $J^{\pi}$=0$^+$ nuclei)
\begin{eqnarray} \label{hadronicratio2}
\Gamma=\frac{1}{\beta_V^{(0)}}\frac{\tilde{F}_{C0} (q)}{F_{C0}(q)}-1
\end{eqnarray}
and it accounts, in particular, for the effects of nuclear isospin
mixing and strangeness content in the PV asymmetry. The ratio between
the WNC and EM form factors can be written as
\begin{equation}
\frac{\widetilde{F}_{C0}(q)}{F_{C0}(q)} =
\frac{\beta_V^{(0)}\langle0||M^{C}_{0\:(T=0)}(q)||0\rangle + \beta_
V^{(1)}\langle0||M^{C}_{0\:(T=1)}(q)||0\rangle}{\langle0||
M^{C}_{0\:(T=0)}(q)||0\rangle +
\langle0||M^{C}_{0\:(T=1)}(q)||0\rangle} \, ,
\label{fft01}
\end{equation}
where the subscript in parenthesis indicates isoscalar ($T=0$) and
isovector ($T=1$) parts.

The Coulomb isoscalar ($T=0$) and isovector ($T=1$) multipole
operators can be written in terms of the contributions of order 0 and
1 in $\eta=p/m_N$ (subscripts (0) and (1)) \cite{Ama96,
Ama96_2,Jes98}:
\begin{equation} \label{operatorM}
M^{C\:M_J}_{J}(q\textbf{x}) = M^{C\:M_J}_{J}(q\textbf{x})_{(0)} +
M^{C\:M_J}_{J}(q\textbf{x})_{(1)} \, .
\end{equation}
The nucleon form factors $G_E$ and $G_M$ are included in the
definitions of the two contributions as follows:
\begin{eqnarray} \label{operatorM2}
M^{C\:M_J}_{J}(q\textbf{x})_{(0)} &=& \frac{\kappa}{\sqrt{\tau}}
G_E(\tau) M^{M_J}_{J}(q\textbf{x})\, ,\\
M^{C\:M_J}_{J}(q\textbf{x})_{(1)} &=&
\kappa\sqrt{\tau}\left[2G_M(\tau)- G_E(\tau)\right]
\Theta^{M_J}_J(q\textbf{x})\, ,
\label{operatorMso}
\end{eqnarray}
where the kinematical factors $\kappa$ and $\tau$ contain the energy
$\omega$ and momentum $q$ transferred to the nucleus by the electron:
\begin{eqnarray}
\kappa=\frac{q}{2m_N}\, , \qquad
\tau=\frac{|Q^2|}{4m_N^2}=\frac{q^2-\omega^2}{4m_N^2}\, ,
\end{eqnarray}
and $G_{E(M)}$ are the electric (magnetic) nucleon form factors
discussed later.  The basic multipole operators $M_J^{M_J}$ and
$\Theta_J^{M_J}$ are respectively the standard zeroth-order
Coulomb operator and the spin-orbit first-order correction, and
are defined as:
\begin{eqnarray}
M^{M_J}_{J}(q\textbf{x}) &=& j_J(q\text{x})Y_J^{M_J}(\hat{\textbf{x}})\, , \\
\Theta^{M_J}_J(q\textbf{x}) &=& -\frac{i}{q^2} \vec{\sigma}\cdot
\left[ \left( \vec{\nabla }M^{M_J}_{J}(q\textbf{x})\times
\vec{\nabla} \right)  \right] \, .
\end{eqnarray}

By using the relations between form factors in isospin space
($T=0,1$) and form factors in charge space (proton:p and neutron:n),
\begin{eqnarray}
M^{C\:M_J}_{J\:(T=0)} &=& M^{C\:M_J}_{J\:(p)}+  M^{C\:M_J}_{J\:(n)} \, , \\
M^{C\:M_J}_{J\:(T=1)} &=& M^{C\:M_J}_{J\:(p}-  M^{C\:M_J}_{J\:(n)} \, , \\
G^{(0)}_E &=& G_{E_{p}}+G_{E_{n}} \, ,\\
G^{(1)}_E &=& G_{E_{p}}-G_{E_{n}} \, ,\\
G^{(0)}_M &=& G_{M_{p}}+G_{M_{n}} \, ,\\
G^{(1)}_M &=& G_{M_{p}}-G_{M_{n}} \, ,\\
\end{eqnarray}
one can write the ratio in Eq.~(\ref{fft01}) as
\begin{equation}
\frac{\widetilde{F}_{C0}(q)}{F_{C0}(q)} =
\frac{\langle0||\widetilde{M}^{C}_{0\:(p)}(q)||0\rangle +
\langle0||\widetilde{M}^{C}_{0\:(n)}(q)||0\rangle}{\langle0||
M^{C}_{0\:(p)}(q)||0\rangle + \langle0||M^{C}_{0\:(n)}(q)||0\rangle}
\, ,
\label{ffpn}
\end{equation}
where the operators with tildes, $\widetilde{M}$, have the same
structure as in Eqs.~(\ref{operatorM}, \ref{operatorM2}), but contain
the WNC nucleon form factors $\widetilde{G}_E $, $\widetilde{G}_M $ to
be defined in the next subsection.

In the present study we are interested in the $J=0$ ($M_J$=0)
multipole. The charge operator $M^0_0$ matrix elements in a spherical
harmonic oscillator (s.h.o.) basis are given by
\begin{equation}\label{M_density}
\langle n'l'j' | M^0_0(q\textbf{x}) | nlj \rangle =
\frac{1}{\sqrt{4\pi}} \: \langle n'l'j' | j_0(qr) | nlj \rangle
\:\delta_{l'l} \: \delta_{j'j} = \frac{1}{\sqrt{4\pi}} \int j_0(qr) \:
R_{nl}(r)R_{n'l}(r) \: r^2 \: dr \, .
\end{equation}
For $J=0$ the spin-orbit operator $\Theta_0^0$ is just proportional to
$\vec{l} \cdot \vec{s}$ and one has in the same s.h.o. basis:
\begin{eqnarray}\label{M_spinorbit}
&& \langle n'l'j' | \Theta^0_0(q\textbf{x}) | nlj \rangle =
-\frac{1}{\sqrt{4\pi}} \:\frac{1}{2}
\left[j(j+1)-l(l+1)-\frac{3}{4}\right] \:\Bigg\langle n'l'j' \Bigg|
\frac{j_1(qr)}{qr} \Bigg| nlj \Bigg\rangle \: \delta_{l'l}
\:\delta_{j'j} = \\\nonumber && -\frac{1}{\sqrt{4\pi}} \:\frac{1}{2}
\left[j(j+1)-l(l+1)-\frac{3}{4}\right] \: \int \frac{j_1(qr)}{qr} \:
R_{nl}(r)R_{n'l}(r) \: r^2 \: dr \, .
\end{eqnarray}
In both expressions $j_{\lambda}(qr)$ is a spherical Bessel function
of order $\lambda$.  The matrix elements of the charge monopole
operators (EM and WNC) between two s.h.o. wave functions are therefore
 \begin{eqnarray}\label{ff_basis_EM}
f_{nn'lj}^{\xi}(q) &\equiv &  \langle n'lj | M^{C\:0}_{0\:(\xi)}
(q\textbf{x}) | nlj \rangle \\\nonumber
&=& \frac{\kappa}{\sqrt{\tau}} G_{E_{\xi}}(\tau) \langle n'lj |
M^0_0(q\textbf{x}) | nlj \rangle + \kappa\sqrt{\tau}\left[2G_{M_{\xi}}
(\tau)-G_{E_{\xi}}(\tau)
\right]  \langle n'lj | \Theta^0_0(q\textbf{x}) | nlj \rangle \,
\end{eqnarray}
\begin{eqnarray}\label{ff_basis_WNC}
\widetilde{f}_{nn'lj}^{\xi}(q) &\equiv &  \langle n'lj |
\widetilde{M}^{C\:0}_{0\:(\xi)}(q\textbf{x}) | nlj \rangle
\\\nonumber &=& \frac{\kappa}{\sqrt{\tau}}
\widetilde{G}_{E_{\xi}}(\tau) \langle n'lj |
M^0_0(q\textbf{x}) | nlj \rangle +
\kappa\sqrt{\tau}\left[2\widetilde{G}_{M_{\xi}}(\tau)-
\widetilde{G}_{E_{\xi}}(\tau)
\right]  \langle n'lj | \Theta^0_0(q\textbf{x}) | nlj \rangle \, ,
\end{eqnarray}
where the proton ($\xi = p$ or $\pi$) and neutron ($\xi = n$ or $\nu$)
form factors $G$ and ${\widetilde{G}}$ will be defined in the next
subsection. Note that one should not confuse $n$ as a superscript on
$f$ and ${\widetilde{f}}$ meaning neutron with $n$ as a subscript
meaning the quantum number labeling the single-particle basis states
--- the context will also make the distinction clear. The expressions
above are used once the s.h.o. expansion of the HF single-particle
wave functions is performed (see below). The total monopole matrix
elements also contain a center-of-mass correction factor to account
for the fact that the HF single-particle wave functions are referred
to the center of the mean-field potential, not to the nuclear center
of mass, as they should be to avoid the spurious movement of the
nucleus as a whole. For a s.h.o. potential, this correction factor
takes the form $f_{CM}=\exp (q^2b^2/4A)\cong \exp (q^2/4A^{2/3})$,
with $q$ the momentum transfer (in fm$^{-1}$), $b$ the oscillator
parameter (in fm) and $A$ the total number of nucleons. This is
usually employed even when not using a s.h.o. basis; the correction
cancels out when one constructs the ratios of form factors appearing
in the asymmetry $\mathcal A$.

Finally, let us conclude this section with a brief discussion of
potential meson-exchange current effects on the PV asymmetry (see also
\cite{Mus94}). Such contributions enter in very different ways in
cross sections and in the asymmetry. In the former they typically
provide effects at the few percent level, whereas for the latter the
situation is not so obvious. First, were there to be no significant
isospin mixing and strangeness contributions, {\it i.e.,} with a
purely isoscalar, no-strangeness ground state, then the MEC effects
would cancel in the ratio that gives the PV asymmetry, and, in fact,
the result would be completely unaffected by any aspect of hadronic
structure. This is an old result \cite{Fei75,Wal77} from the earliest
studies of elastic PV electron scattering from nuclei. Secondly, with
isospin mixing the situation is more complicated, since now there are
both isoscalar and isovector electromagnetic current matrix elements
involved and MEC effects are different for the two.  However, two
arguments suggest that these contributions are not significant for the
present study at low momentum transfers: the multipoles involved are
C0 (isoscalar and isovector) where MEC effects, being dominantly
transverse, are suppressed anyway and the effect is one of modifying
the predicted isovector effect from occurring as a purely one-body
matrix element to one where both one-body and typically few percent
two-body contributions occur. This would modify the isospin-mixing PV
results by a multiplicative (not additive) factor of typically a few
percent, which is not significant for the present study. Thirdly, with
strangeness also present, now in the nucleon itself and also in MEC
contributions, additional effects may occur. A study was done
\cite{Mus94a} for the important case of $^4$He and, based on the
results presented there, it may be that at higher values of $q$
(beyond roughly 3 fm$^{-1}$) such effects could become significant. On
the other hand, in the cited study such effects were shown to be
negligible for the range of momentum transfers of interest in the
present work and accordingly here we also neglect any contributions of
this type.

\subsection{Nucleon form factors and strangeness}

The electric nucleon form factor can be expressed either in terms of
protons and neutrons or in terms of isoscalar and isovector parts,
\begin{equation}
G_E = G_{E_{p}}\frac{1}{2}(1+\tau_3) + G_{E_{n}}\frac{1}{2}(1-\tau_3)
= \frac{1}{2}(G_E^{(0)} + \tau_3 G_E^{(1)})\, ,
\end{equation}
where $\tau_3$ is 1 for protons and $-1$ for neutrons. In analogy, in
the Standard Model, the WNC nucleon form factor is given by
\begin{equation}
\widetilde{G}_E=\frac{1}{2}(\beta_V^{(0)} G_E^{(0)} + \beta_V^{(s)}
G_E^{(s)} + \tau_3 \beta_V^{(1)} G_E^{(1)})\, ,
\end{equation}
where the strangeness contribution to the form factor is
isoscalar. This can also be written in terms of proton and neutron
form factors,
\begin{equation}
\widetilde{G}_E = \widetilde{G}_{E_{p}}\frac{1}{2}(1+\tau_3) +
\widetilde{G}_{E_{n}} \frac{1}{2}(1-\tau_3)\, ,
\end{equation}
where
\begin{eqnarray} \label{WNCff}
\widetilde{G}_{E_{p}} = \beta_V^{p} G_{E_{p}} + \beta_V^{n} G_{E_{n}}
+ \frac{1}{2} \beta_V^{(s)} G_{E}^{(s)} \, , \\ \widetilde{G}_{E_{n}}
= \beta_V^{n} G_{E_{p}} + \beta_V^{p} G_{E_{n}} + \frac{1}{2}
\beta_V^{(s)} G_{E}^{(s)} \, .
\end{eqnarray}
A similar result applies to the nucleon magnetic form factors by
simply substituting G$_E$ by G$_M$ in the previous expressions.  The
nucleon form factors $G_E$ and $G_M$, from which $\widetilde{G}_E$ and
$\widetilde{G}_M$ are obtained, have been computed using the
parametrization by H\"ohler \cite{Hoh76}.

Within the Standard Model we have,
\begin{eqnarray}
\beta_V^{(0)} &=& \beta_V^{p} + \beta_V^{n} = -2 \sin ^2 \theta_W
=-0.46 \\ \beta_V^{(1)} &=& \beta_V^{p} - \beta_V^{n} = 1 -2 \sin ^2
\theta_W =0.54 \\ \beta_V^{p} &=& \frac{1}{2} \left( \beta_V^{(0)}+
\beta_V^{(1)}\right)= 0.04 \\ \beta_V^{n} &=& \frac{1}{2} \left(
\beta_V^{(0)}- \beta_V^{(1)}\right)= -0.5 \\ \beta_V^{(s)} &=& -1
\end{eqnarray}
The strangeness form factor $ G_{E}^{(s)}$ has been parametrized
according to
\begin{equation}
G_{E}^{(s)} = \rho_s \tau G_D^V \xi_E^{(s)}, \qquad G_{M}^{(s)} =
\mu_s G_D^V \, ,
\label{str_para_1}
\end{equation}
(see, for instance, \cite{Mus94}) with
\begin{equation}
G_D^V = (1+4.97 \tau)^{-2}, \quad \xi_E^{(s)}=(1+ 5.6 \tau)^{-1} \, .
\label{str_para_2}
\end{equation}
The parameters $\rho_s$ and $\mu_s$ are constrained by PV electron
scattering measurements on hydrogen, deuterium and helium-4; the
values chosen as representative are discussed below and in Sect. III.

\subsection{Contributions to the PV asymmetry}

The strangeness term in the WNC form factors in Eq.~(\ref{WNCff}) can
be considered separately, giving rise to a decomposition of the PV
asymmetry deviation, $\Gamma = \Gamma^{(I)} + \Gamma^{(s)}$, where the
isospin mixing term $\Gamma^{(I)}$ is proportional to
$\beta^{(1)}_V/\beta^{(0)}_V$ and $\Gamma^{(s)}$ is proportional to
$\beta^{(s)}_V/\beta^{(0)}_V$. The isospin mixing piece is computed
considering $G_{E}^{(s)}=0$ and $G_{M}^{(s)}=0$ in the WNC form
factors, whereas the strangeness term is computed considering
$G_{E_{p}}=0$, $G_{E_{n}}=0$, $G_{M_{p}}=0$ and $G_{M_{n}}=0$ in the
WNC form factors.

We evaluate the strangeness contribution neglecting the small
differences between neutron and proton densities, assuming
G$_{E_{n}}$=0, and neglecting the small spin-orbit contribution so
only $G_E^{(s)}$ enters and not $G_M^{(s)}$. This gives
\begin{equation}\label{gamma_s}
\Gamma^{(s)} =
\frac{\beta_V^{(s)}}{\beta_V^{(0)}}\frac{G_E^{(s)}}{G_E^{(0)}} \, .
\end{equation}

Once we have introduced the nucleon form factors, it is interesting to
notice that if one neglects the spin-orbit correction to the Coulomb
operator, Eq.~(\ref{operatorMso}), one obtains the more intuitive
expression
\begin{equation}
\frac{\widetilde{F}_{C0}(q)}{F_{C0}(q)}=
\frac{ \widetilde{G}_{E_{p}}
\langle0||M_{0\:(p)}(q)||0\rangle +
\widetilde{G}_{E_{n}}
\langle0||M_{0\:(n)}(q)||0\rangle}
{G_{E_{p}} \langle0||
M_{0\:(p)}(q)||0\rangle + G_{E_{n}}\langle0||M_{0\:(n)}(q)||0\rangle} \, ,
\end{equation}
with
\begin{equation}
\langle0||M_{0\:(\xi)}(q)||0\rangle  \sim
\int j_0(qr) \rho_{(\xi)}(r) r^2 dr \, ,
\end{equation}
where $\rho_{(\xi)}$ are the ground-state radial densities for protons
and neutrons with $\xi$ as above. If one further neglects the electric
neutron form factor $G_{En}$ and the strangeness form factors
$G_E^{(s)}$, one arrives at the simple expression
\begin{equation}
\frac{\widetilde{F}_{C0}(q)}{F_{C0}(q)}=
\beta_V^{p}+\beta_V^{n} \frac{ \langle0||M_{0\:(n)}(q)||0\rangle }
{ \langle0||M_{0\:(p)}(q)||0\rangle } \, ,
\end{equation}
or equivalently
\begin{equation}
\Gamma=\frac{\beta_V^{n}}{\beta_V^{(0)}} \left(
\frac{\langle0||M_{0\:(n)}(q)||0\rangle -
\langle0||M_{0\:(p)}(q)||0\rangle}{\langle0||M_{0\:(p)}(q)||0\rangle}
\right)
\end{equation}
which are similar to those used in DDS.

\subsection{Kinematics and the figure-of-merit}

As a general consideration, one faces a compromise between optimizing
the PV signal ({\it i.e.,} deviations in the asymmetry) and the
figure-of-merit, discussed below. Since the asymmetry generally
increases with $|Q^2|$, while the cross section decreases, this leads
to the search for a well defined region of optimal kinematics. To
measure properly effects in the PV asymmetry due to isospin mixing,
the deviation should be greater than or of the order of a few percent
of the reference value ${\cal{A}}^0$ given in
Eq.~(\ref{referencevalue}). This condition will be fulfilled in
general for some intervals of $q$, but not for others. At the same
time the relative error of the asymmetry should be kept as small as
possible. This relative error depends on technical characteristics of
the experimental setting (detector solid angle, beam luminosity and
running time) gathered in $X_0$, as well as on intrinsic properties of
the target, the projectile and their kinematics gathered in the so
called figure-of-merit (FOM) $\cal{F}$ (see \cite{Mus94}):
\begin{equation}
\frac{\delta {\mathcal A}}{{\mathcal A}}=\frac{1}{\sqrt{{\mathcal F}\: X_0}}\, .
\end{equation}
Clearly, in a study of the present type where the focus is placed on
theoretical aspects of the hadronic and nuclear many-body physics in
the problem, it is not appropriate to go into further detail on
specific experimental conditions, {\it i.e.,} on the specifics of
$X_0$ (see, however, \cite{Mus94} for some general discussions of
experimental aspects of PV electron scattering).  Presenting the FOM,
as we do in the following section, provides a sufficient measure of
the ``doability'' of PV elastic scattering asymmetry measurements
whose goal is to explore isospin mixing in the nuclear ground state.
In particular, the FOM is proportional to the asymmetry itself squared
and to the differential cross section of the scattering (essentially
the parity-conserving cross section):
\begin{eqnarray}
{\mathcal F}=\frac{d \sigma}{d \Omega} {\mathcal A}^2 \, ,
\end{eqnarray}
and therefore shows a diffraction pattern, but with the same
negative-slope trend as the differential cross section when the
momentum transfer increases, which favors low momentum transfer
experiments.

\subsection{HF+BCS ground-state density}

To generate the ground-state wave function in the HF+BCS approximation
we use a Skyrme type density-dependent nucleon-nucleon interaction
(SLy4 force \cite{sly4}), allowing for axially-symmetric
deformation. First, the HF equations are solved to generate the
self-consistent HF mean field, and pairing correlations are taken into
account at each iteration solving the BCS equations to generate the
single quasiparticle wave functions.  The deformed HF+BCS calculation
gives a set of single-particle levels, occupation numbers, and wave
functions $\Phi_{HF}^i$. The latter are expanded in a deformed
harmonic oscillator basis $\vartheta_{n_{\rho}n_{z}\Lambda\Sigma}$
which we also expand in a s.h.o. basis $\phi_{nlj(m_j)}$ to facilitate
the comparison with previous work based on the shell-model approach,
\begin{equation}
\Phi_{HF}^i(\vec{r}) = \sum_{n_{\rho}n_{z}\Lambda\Sigma} k_{n_{\rho}n_{z}
\Lambda\Sigma}^{i} \: \vartheta_{n_{\rho}n_{z}\Lambda\Sigma}(\vec{r}) =
\sum_{nlj} c_{nlj}^{i} \: \phi_{nljm_j}(\vec{r}) \, ,
\end{equation}
where a truncation of 11 major shells $N$ has been used in both the
deformed ($N=2n_{\rho}+n_z+\Lambda$) and the spherical ($N=2n+l$)
expansions. Each HF single-particle state $i$ has a parity and an
angular momentum projection $K^{\pi}$ which are shared by any of the
basis wave functions taking part in its expansion
($\pi=(-1)^l=(-1)^{n_z+\Lambda}$, $K=m_j=\Lambda+\Sigma$). The HF+BCS
outputs give for every state $i$ the energy, the occupation
probability $v^2_i$ and the coefficients $k^i$. The coefficients in
the s.h.o. basis expansion $ c_{nlj}^{i}$ are obtained as
\begin{equation}
 c_{nlj}^{i} = \sum_{n_{\rho}n_{z}\Lambda\Sigma}  k_{n_{\rho}n_{z}
\Lambda\Sigma}^{i}  C^{n_{\rho}n_{z}\Lambda\Sigma }_{Nljm} \, ,
\end{equation}
with
\begin{equation}
C^{n_{\rho}n_{z}\Lambda\Sigma }_{Nljm} = \langle l \:\frac12\: \Lambda\: \Sigma\:
|\: j\:m \rangle (-1)^{\Lambda} \sqrt{\frac{(2l+1)(l-\Lambda)!}
{2(l+\Lambda)!}} \int r^2 dr R_{nl}(r) F(r)
\end{equation}
and
\begin{equation}
F(r)=2\int_0^1 dt P^{\Lambda}_l(t)
\psi^{\Lambda}_{n_{\rho}}(r^2(1-t^2)) \psi_{n_z}(rt)\, ,
\end{equation}
where the $P^{\Lambda}_l$ are associated Legendre polynomials and the
functions $\psi$ are defined in terms of Hermite and generalized
Laguerre polynomials that contain, respectively, the cylindrical
$z=rt$ and $\rho^2=r^2(1-t^2)$ dependence of the eigenstates
$\vartheta_{n_{\rho}n_{z}\Lambda\Sigma}(\vec{r})$ of the deformed
harmonic oscillator potential \cite{vautherin, Moy91}.

The contribution of each pair of s.h.o. states to the total charge
monopole form factors (EM or WNC) can be analyzed in terms of two
pieces as follows:
\begin{equation}
\langle0||M^{C}_{0\:(\xi)}(q)||0\rangle = \sum_{nn'lj}
\: f_{nn'lj}^{\xi}(q) \: \rho_{nn'lj}^{\xi} \, ,
\end{equation}
\begin{equation}
\langle0||\widetilde{M}^{C}_{0\:(\xi)}(q)||0\rangle = \sum_{nn'lj}
\: \widetilde{f}_{nn'lj}^{\xi}(q) \: \rho_{nn'lj}^{\xi} \, ,
\end{equation}
applicable to protons and neutrons ($\xi = p,\ n$) separately. The
first factor in the above equations is the matrix element of the
charge monopole operator (EM or WNC) between two s.h.o. states as
defined in Eq.~(\ref{ff_basis_EM}) or (\ref{ff_basis_WNC}), which is
momentum-dependent and whose $j$-dependence is only due to the
spin-orbit correction (Eq.~(\ref{M_spinorbit})). This matrix element
vanishes unless both basis functions have the same $lj$ quantum
numbers. The nuclear ground-state structure information is contained
in the second factor, which is the spherical part of the HF+BCS
ground-state density matrix (for protons or neutrons) in the
s.h.o. basis:
\begin{equation} \label{density}
\rho_{nn'lj}=\sum_{i} \: 2v_i^2 \: c_{nlj}^{i} c_{n'lj}^{i} \:\:.
\end{equation}
It contains the coefficients of pairs of spherical harmonic oscillator
components differing at most in the radial quantum number $n$, as well
as the occupation probabilities $v_i^2$ of each of the HF
single-particle states $i$. These contributions are added up so that
the calculated density refers to the whole HF+BCS ground state of the
nucleus. The spherical density matrix elements contain information on
the nuclear ground-state structure of the target isotopes. We note
that the total HF+BCS density matrix may also contain a non-spherical
part, with matrix elements $\rho_{nn'll'jj'}$, which for deformed
nuclei is nonzero and which does not contribute to the charge monopole
form factors.  This ensures that only the spherically symmetric
($J=0$) part of the neutron and proton densities contributes and that
there is no need to make any further angular momentum projection in
the deformed nuclei~\cite{Moy86, Zar77}.  The analysis of the
quantities defined above is especially appropriate when comparing
results of our HF+BCS calculations with former shell-model results,
since each quasiparticle state can always be expressed as a
combination of s.h.o. basis states with different radial quantum
numbers $n$.  This fact allows for nonzero spherical density matrix
elements $\rho_{nn'lj}$ with different $n$ and $n'$, off-diagonal in
the spherical harmonic oscillator basis.

When the Coulomb interaction is included in the generation of the
self-consistent mean field, the spherical density matrix in
Eq.~(\ref{density}) is slightly different for protons and
neutrons. This amounts to saying that the self-consistent ground-state
mean field is not a pure $T=0$ isospin state, but contains isospin
admixtures, mainly $T=1$ \cite{Alv05}, which contribute to the PV
$\Delta T=1$ amplitude. In other words, from the proton and neutron
densities $\rho_{nn'lj}^{\xi}$, $\xi = p,\ n$, one may construct an
isoscalar ground-state density
$\rho_{nn'lj}^{T=0}=\rho_{nn'lj}^{\xi=p}+\rho_{nn'lj}^{\xi=n}$ and an
isovector ground-state density
$\rho_{nn'lj}^{T=1}=\rho_{nn'lj}^{\xi=p}-\rho_{nn'lj}^{\xi=n}$ that
contribute respectively to the isoscalar and isovector parts of the PV
amplitude.

\subsection{Coulomb distortion effects}

Another aspect that must be addressed is the Coulomb distortion of the
incoming and outgoing electron wave functions. While we are dealing
with relatively light nuclei for which Coulomb distortion effects are
generally small (at least where the FOM is significant), it is
important to perform calculations which fully take into account
Coulomb distortion to obtain realistic predictions that can serve as a
reference for future experiments. It is also interesting to compare
these results with the PWBA calculations described in previous
sections.  To this end, we follow the standard treatment of Coulomb
distortion for elastic PV electron scattering within a partial wave
formalism.  We solve the Dirac equation for massless electrons in the
Coulomb potential generated by the nucleus, closely following the work
of refs. \cite{Ruf82,Hor98,Ant05} to obtain the distorted wave (DW)
results.

\begin{figure*}
\centering
\includegraphics[width=130mm]{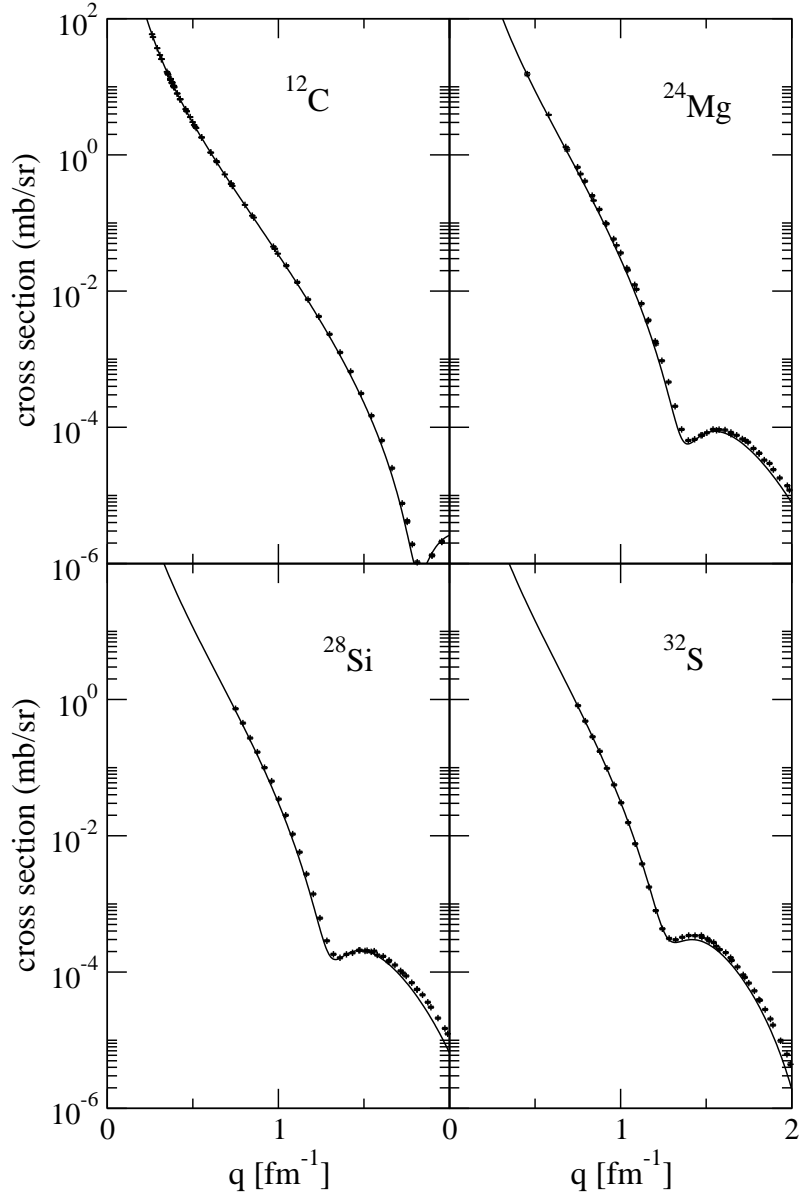}
\caption[]{Experimental data with error bars and theoretical (solid
  line) electron-nucleus cross section. Ground-state nuclear
  densities are obtained from a HF(SLy4)+BCS($\Delta_{\pi,\nu}$= 1
  MeV) distorted wave calculation for 400 MeV electrons.}
\label{xs}
\end{figure*}

In Fig.~\ref{xs} we compare our unpolarized elastic electron
scattering cross sections with experimental data for $^{12}$C
\cite{Sick70,Jansen72,Fey73,Cardman80,Reuter81}, for $^{24}$Mg
\cite{Li74,Lees76}, for $^{28}$Si \cite{Li71a} and for $^{32}$S
\cite{Li71a}. For ease of presentation, the cross sections have been
transported to a common energy of 400 MeV. Our calculations of cross
sections have been obtained from the HF+BCS ground-state densities as
described in the previous section. In obtaining these results, no
experimental information on, for instance, charge radii, has been
employed or fit. In spite of this fact, very good agreement is found
up to a scattering angle of 60$^\circ$, {\it i.e.,} a transfer
momentum up to around 2 fm$^{-1}$, which is the region of interest
according to the results on FOM to be shown below.

\begin{figure*}
\centering
\includegraphics[angle=-90,width=130mm]{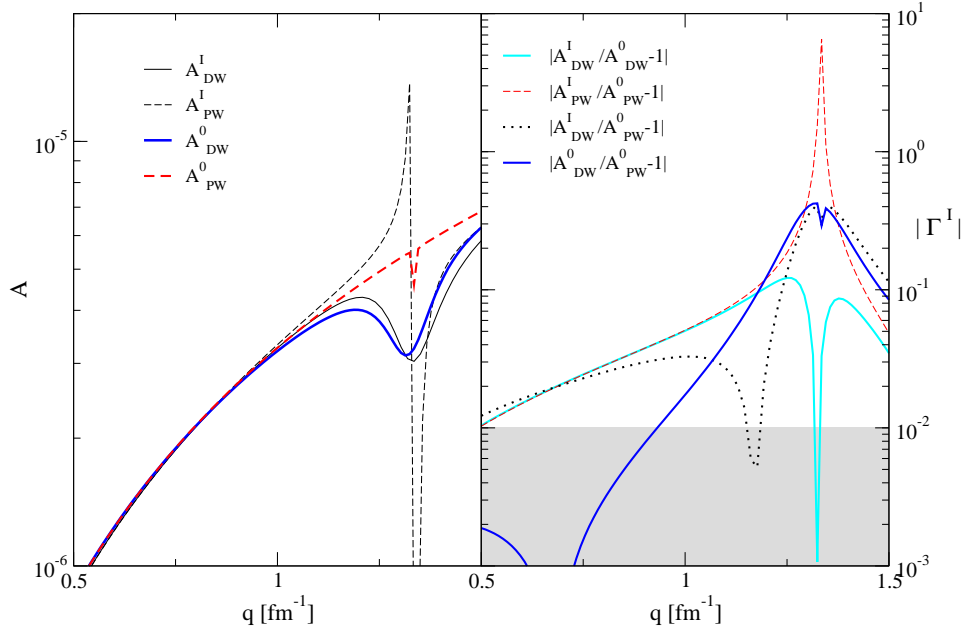}
\caption[]{(Color online) Left-hand panel: PV asymmetry for $^{28}$Si
  allowing for nuclear isospin mixing in PWBA and in the fully
  distorted calculation (PW$^{\text{I}}$ and DW$^{\text{I}}$) and in
  the non-isospin-mixing case by constraining $\rho^{p} =
  \rho^{n}$ (PW$^0$ and DW$^0$).  Right-hand panel: Asymmetry
  deviations for $^{28}$Si: due to pure isospin mixing effects in DW
  (solid line) and in PWBA (dashed line), due to isospin mixing and
  distortion effects together (dotted line) and due to distortion
  effects only (thick solid line). }
\label{asym_gamma_DW_PW}
\end{figure*}

To consider the distorted-wave correction in the PV asymmetries,
Eq.~(\ref{asymmetry_sigmas}) has been used, together with
Eq.~(\ref{asymmetryratio}) for the asymmetry deviations. The effects
of Coulomb distortion and nuclear isospin mixing can be analyzed
separately by considering different ingredients in the calculation of
the asymmetries in Eq.~(\ref{asymmetryratio}). In
Fig.~\ref{asym_gamma_DW_PW} we show the typical distortion effects for
1 GeV electrons in the case of $^{28}$Si. The left-hand panel shows
isospin-mixing (superscript I) asymmetries in PWBA and when
distortions of the electron wave function are taken into account,
where the smoothing of the PWBA divergences appears as the most
obvious effect of distortion. One must be aware of the fact that
without isospin effects and distortion of the electrons, the WNC and
EM form factors in Eq.~(\ref{hadronicratio}) are exactly proportional,
and thus the asymmetry follows a very simple $|Q^2|$ dependence (shown
by the long dashed line in the left-hand panel of the figure), even at
the zeros of the form factor corresponding to the diffraction
minima. When isospin effects are considered, the diffraction minima
for the WNC and EM form factors occur at slightly different values of
$q$ and thus the asymmetry shows extreme variations at the approximate
locations of these diffraction minima, as can be seen in the left-hand
panel of Fig.~\ref{asym_gamma_DW_PW} (dashed curve). The main effect
of the distortion of the electron waves for these light systems is to
fill in or to smooth out the diffraction minima, thus severely
reducing the amplitude changes of the asymmetry at these diffraction
minima, as can also be seen (solid curves) in
Fig.~\ref{asym_gamma_DW_PW}. Furthermore, the distorting potential
introduces asymmetry deviations even when no isospin mixing effects
are present in the calculation of the nuclear structure.

In summary, when nuclear isospin mixing is not considered, {\it i.e.,}
fixing the same proton and neutron densities, the PWBA result shows
simply a $|Q^2|$ behaviour, while the results that take into account
the distortion of the electron wave function deviate smoothly from
this with a dip where the diffraction minimum occurs. When isospin
mixing is included, the asymmetry is increased with respect to the
non-isospin-mixing case as $q$ approaches the value at the diffraction
minima, and it is reduced after the diffraction minima. This behavior
is seen both in the PWBA and in the DW calculations.

In the same Fig.~\ref{asym_gamma_DW_PW}, the plots in the right-hand
panel show asymmetry deviations $\Gamma^{\text{I}}$ due to the nuclear
isospin mixing obtained replacing ${\mathcal A}/{\mathcal A}^0$ in
Eq.~(\ref{asymmetryratio}) by the ratios ${\mathcal
A}^{\text{I}}_{PW}/{\mathcal A}^0_{PW}$ and ${\mathcal
A}^{\text{I}}_{DW}/{\mathcal A}^0_{DW}$. In addition we show the pure
effect of distortion (ignoring nuclear isospin mixing) obtained from
the ratio ${\mathcal A}^0_{DW}/{\mathcal A}^0_{PW}$ and the combined
effect of distortion and isospin mixing yielded by using the ratio
${\mathcal A}^{\text{I}}_{DW}/{\mathcal A}^0_{PW}$. As one can see in
the figure, the effect of distortion is to smooth out the divergences
appearing at the position of diffraction minima in PWBA, but anywhere
else one sees that the deviations in the asymmetry introduced by
isospin mixing effects are very similar both in PWBA and when the
distortions are fully taken into account. Analogous results are
obtained for $^{12}$C, $^{24}$Mg and $^{32}$S. One should notice that,
since we are plotting the absolute value of the asymmetry deviation on
a logarithmic scale, zeros of this function appear as downward
divergences which remain in the DW calculation.

In the next section we focus the discussion on the results
corresponding to the asymmetry deviation. This is defined as the ratio
whose numerator is the difference between the asymmetry with isospin
mixing and without, and whose denominator is simply the asymmetry with
no isospin mixing, where all asymmetries are from full distorted-wave
calculations. That is, one has
\begin{eqnarray} \label{asymmetryratio_DW}
\Gamma^{\text{I}}_{DW}\equiv\frac{{\mathcal
A}^{\text{I}}_{DW}}{{\mathcal A}^0_{DW}}-1 \, .
\end{eqnarray}
This deviation may be directly compared with experiment.

It can be seen that this generalized $\Gamma^{\text{I}}_{DW}$ yields
results that are similar to the PWBA prediction, except for the
regions deep in the minima of the cross section. Taking into account
that ${\mathcal A}^0_{DW}$ can be computed accurately in a
model-independent way (tables can be produced for each nucleus and
different electron energies without difficulty), and the only inputs
needed are the (experimental) charge distributions, which are well
known for the nuclei studied here, we conclude that the distortion
effects will not prevent the determination of isospin mixing effects,
provided that the data are compared with a full DW
calculation. Indeed, the organization of the asymmetry data according
to $\Gamma^{\text{I}}_{DW}$ introduced in the equation above may be
enough to permit a direct comparison with the simple PWBA predictions
for $\Gamma$.

It is worth pointing out that the filling of the minima of the cross
section is important for the determination of the asymmetry and of the
deviations from the non-isospin-mixing prediction, as it is precisely
the region near the minima where the isospin mixing effects on the
asymmetry are also more evident. The filling of the minima of the form
factor and cross section induced by Coulomb distortion will help to
make these details of the asymmetry more easily measurable, as the
cross section will be much larger in the region of the minima than
predicted with the PWBA calculation.

\section{Results}

\subsection{Exploration of basic assumptions}

Four even-even, N=Z isotopes have been studied in this work, namely
$^{12}$C, $^{24}$Mg, $^{28}$Si and $^{32}$S, which for each element
are the most abundant isotopes (79$\%$ abundance for $^{24}$Mg, and
higher than 90$\%$ for the other isotopes).  Furthermore, all targets
have reasonably large excitation energies of the first (2$^+$) excited
state, and all but sulfur are suitable in elemental form for
high-current electron scattering experiments, as required for a
measurement of the PV asymmetry. In our calculations the ground-state
deformation is obtained self-consistently. For each nucleus the
calculations have been performed using the optimum values of the
axially-symmetric harmonic oscillator parameters $b$ and $q$
\cite{vautherin}, which define the average length and the axes ratio
of the oscillator well. The self-consistent intrinsic proton
quadrupole moments are shown in Table \ref{tabledef} for three
different sets of pairing gaps: $\Delta_{\pi,\nu}$ = 0 MeV (absence of
pairing); $\Delta_{\pi,\nu}$ = 1 MeV; and $\Delta_{\pi,\nu}$ as
obtained from experimental mass differences \cite{Aud03} of
neighboring nuclei through a symmetric five-term formula.  These
values of the pairing gaps are practically the same for protons and
neutrons, namely 4.5 MeV for $^{12}$C, 3.13 MeV for $^{24}$Mg, 2.88
MeV for $^{28}$Si, and 2.17 MeV for $^{32}$S.  Since these isotopes
are relatively light, these pairing gaps are too large because shell
effects are not taken into account in the mass formula.  The larger
pairing gaps tend to give smaller self-consistent deformations. For
$^{12}$C and $^{32}$S we obtain spherical equilibrium shapes in
agreement with their vibrational experimental spectra. Table
\ref{tabledef} also includes experimental intrinsic charge quadrupole
moments \cite{Rag89}. For $^{24}$Mg and $^{28}$Si, whose experimental
spectra show a rotational band that confirms their deformed shape, the
self-consistent values of the quadrupole moment are in better
agreement with experiment for the smaller pairing gap values. For this
reason, in what follows we focus on results obtained for
$\Delta_{\pi,\nu}$ = 1 MeV.  Values of $Q_{0\:p}= -21$ fm$^2$ and
$Q_{0\:p}$= 52.15 fm$^2$ are reported for $^{12}$C and $^{32}$S
respectively in \cite{Rag89}.  However, in these two isotopes a
ground-state rotational band is by no means recognizable \cite{Fir99},
implying that their intrinsic quadrupole ground-state values should be
zero and that the ones appearing in \cite{Rag89} do actually
correspond to their vibrational 2$^+$ states. The interpretation of
$^{12}$C in terms of a 3$\alpha$ structure is gaining more and more
experimental support~\cite{Fyn05}. Obviously this picture is not
reducible to a single determinantal function. However, we note the
fact that the HF+BCS energy which we obtain as a function of
quadrupole deformation shows a shallow minimum around $Q$ = 0 for
$^{12}$C, and this is consistent with that picture. In any case, the
final results for PV asymmetries are not very sensitive to the nuclear
deformation. This was tested by computing the PV asymmetry for two
different shapes of $^{32}$S, spherical ($Q_{0p} \sim$ 0 fm$^2$, which
is the self-consistent deformation) and prolate ($Q_{0p} \sim$ 50
fm$^2$, which has been fixed by means of a quadrupole constraint in
the HF calculation). The results for the two cases were found to be
extremely similar, which shows that our predictions are quite
independent of the nuclear shape as long as the corresponding
self-consistent densities are properly used in the calculations.

In the figures to follow we present the results of isospin mixing for
the four isotopes under study, using a SLy4 HF+BCS mean field with
$\Delta_{\pi,\nu}$=1 MeV. For comparison, although this is not the
primary focus of the present study, we also show results with/without
strangeness present, and accordingly a few words are in order
concerning what should be regarded to be a sensible value for
$\Gamma^{\text{s}}$. The experimental value for $\rho_s$ in
Eq. (\ref{str_para_1}) is still evolving, given the appearance of new
experiments, more detailed analyses of the combined experimental
results (see {\it e.g.,} \cite{Leinweber06}) and considerations of
complications such as isospin mixing in the nucleon \cite{Kubis06} and
$^4$He \cite{Don89,Rama94,Viviani07} and the role of two-photon
exchange \cite{Zhou07}. Presently, from PV electron scattering
measurements involving $^{1,2}$H and $^4$He, the value of $\rho_s$ is
consistent with zero, and accordingly in the figures we show
isospin-mixing results computed with $\rho_s=0$. The strongest
constraint on electric strangeness to date comes from the HAPPEX-He
experiment at JLab \cite{Acha07} which yielded the result
$G_E^{(s)}=0.002\pm 0.014 (stat) \pm 0.007 (syst)$ at $q=$1.4
fm$^{-1}$ (corresponding to $Q^2=0.077$ (GeV/c)$^2$), which translates
into a range for the electric strangeness parameter used in our
parametrization (see Eqs. (\ref{str_para_1}) and (\ref{str_para_2}))
of $-1.2 \lesssim \rho_s \lesssim +1.4$; to be conservative we have
simply added statistical and systematic errors here.  Accordingly, in
the present study, when referencing our new results to existing
estimates of electric strangeness we have adopted the two limit values
$\rho_s=\pm1.5$ as a rough measure when strangeness is present, and
$\rho_s=0$ when it is not --- all three of them are essentially
consistent with existing knowledge on electric strangeness. Of course,
if (when) the amount of electric strangeness is better known from PV
electron scattering studies of $^{1,2}$H and $^4$He the present
analysis can also easily be updated, both via more refined
parametrizations of $G_E^{(s)}$ and in concert with the isospin mixing
effects that provide the main focus of the present work. With
reference to the results shown below, one should note that the
HAPPEX-He PV measurements have been performed in comparable low-$q$
kinematical regions.

For direct comparison with $\Gamma^{\text{I}}$ we use the
$\Gamma^{\text{s}}$ of Eq.~(\ref{gamma_s}). Note that when $\rho_s$ is
negative the effects of isospin mixing and strangeness tend to cancel
in the low-$q$ region ($q \lesssim$ 1 fm$^{-1}$), whereas the opposite
would be true were $\rho_s$ to be positive, which is not ruled out at
present. Clearly, in the case where no significant electric
strangeness is found in the nucleon the curves with isospin mixing and
$\rho_s =0$ are the appropriate ones.  It should also be noted that
the strangeness contribution is not very sensitive to details of the
nuclear structure (in particular, the approximation of
Eq.~(\ref{gamma_s}), which is the one plotted in the graphs, is
completely independent of the nuclear structure), whereas the isospin
contribution does depend on the details of the nuclear structure. For
magnetic strangeness (see Eqs. (\ref{str_para_1}) and
(\ref{str_para_2})) following \cite{Mus94} we have assumed that $\mu_s
= -0.31$, although the effects in this case are very small and
choosing any value that is consistent with present experimental
knowledge would lead to negligible differences.

We now begin our discussions of some of the main results of our work,
focusing here on an exploration of the basic assumptions made in the
present approach, using the case of $^{28}$Si to illustrate the
various effects under study.  In Fig. \ref{gamma_iso_strange_si28} the
left-hand panel shows the total PV asymmetry deviation $\Gamma$ using
the form factor ratio in Eq. (\ref{ffpn}), together with the isospin
mixing contribution $\Gamma^{\text{I}}$ and the strangeness
contribution $\Gamma^{\text{s}}$ discussed above (as well as below,
where the full set of nuclei are inter-related). Discussion of the
right-hand panel, which shows the FOM as well as a comparison of
plane- and distorted-wave results, is also postponed to later where
results for the full set of nuclei are presented. At that point the
``doability'' of measurements of elastic PV electron scattering is
discussed in a little more detail.  However, before proceeding with
those general results, let us comment on some of the specifics, using
the case of $^{28}$Si as prototypical.

First, the effect of using different values for the pairing gap in
obtaining the asymmetry deviation is shown in
Fig.~\ref{gamma_pairing_si28}, where one can see that the results
change very little, even for the larger pairing gap values, as long as
the proton and neutron gap values are the same. In conclusion, our
results on the PV asymmetry, as shown in the figure, are very stable
against changes in the proton and neutron pairing gaps, provided both
have the same value. If the proton and neutron pairing gaps were to be
notably different, as in one of the examples of
Fig.~\ref{gamma_pairing_si28}, this would lead to a very different
structure of the proton and neutron distributions which translates
into a different shape for the PV asymmetry deviation.

Secondly, let us discuss the sensitivity of the results to the use of
different Skyrme interactions. In Fig. \ref{gamma_forces_si28} results
for $^{28}$Si are shown for three different choices, SLy4, Sk3 and
SG2. SLy4 force \cite{sly4} is an example of a recent parametrization
and Sk3 \cite{sk3} is an example of a simple and old force. SG2
\cite{sg2} has been also widely used in the literature and provides a
good description of bulk and spin-isospin nuclear properties as well.
Clearly in the low-$q$ region of interest in the present work there is
very little sensitivity to the choice of force. The small spread in
$\Gamma$ seen in the figure may be taken as a sort of theory
uncertainty with which other competing effects can be compared.
 
Thirdly, consider the spin-orbit contribution which was introduced as
a correction in Eq.~(\ref{operatorM}). While expected to be a small
correction, it is important to make sure that this contribution does
not confuse the interpretation of the PV asymmetry in terms of isospin
mixing.  Typically when treating parity-conserving elastic electron
scattering the spin-orbit contribution is dominated by isoscalar
effects and is known to provide a small correction. However, since the
isospin mixing involves a delicate interplay between isoscalar and
isovector matrix elements and since the isovector spin-orbit
contribution in particular may be sizable (see
Eq.~(\ref{operatorMso})): the isovector/isoscalar form factor ratio
there is $[2 G_M^{(1)}-G_E^{(1)}]/[2 G_M^{(0)}-G_E^{(0)}] \cong 11$),
this contribution has also been included in the present work for
completeness. Results obtained for the PV asymmetry deviation in
$^{28}$Si performed with and without the spin-orbit correction are
almost indistinguishable for momentum transfers lower than about 3.5
fm$^{-1}$, except near the diffraction minima where the spin-orbit
contribution becomes (fractionally) significant and in the dips where
$|\Gamma|$ is small anyway. One concludes that, at least in those
regions away from the peaks seen in $|\Gamma|$, one should not expect
the spin-orbit contributions to lead to confusion when attempting to
extract the isospin-mixing behaviour of the PV asymmetry.

Fourthly, a comparison between previous shell-model calculations and
our results is now in order. By examining, for instance, Figs.~3 or 7
in DDS together with the present results we immediately see that the
effects of isospin mixing are considerably larger here. In the
shell-model calculation of DDS, the value of the diagonal density
matrix elements for those states below the active shell was the
maximum possible occupation of the states in the isoscalar case and
zero in the isovector case. For the states in the active shell, only
the sum of the density matrix elements was fixed, but not the
individual values. All the off-diagonal density matrix elements
vanished in the shell-model calculation, but not in the HF calculation
for the reasons mentioned above. These off-diagonal matrix elements
are the main contributors to the total isovector form factors obtained
in the HF case. In particular, it is of interest to trace back which
contributions of the isovector spherical density matrix elements
(Eq.~(\ref{density})) contain the largest major shell mixings. To
illustrate this analysis we show in Table \ref{multipolar} the
spherical isovector density matrix elements (diagonal and
off-diagonal) for different ($l,j$) contributions of $^{28}$Si. It is
found that the off-diagonal isovector density between the harmonic
oscillator levels 0d$_{5/2}$ and 1d$_{5/2}$, {\it i.e.,}
$\rho^{T=1}_{012\frac{5}{2}} =
\rho^{\xi=p}_{012\frac{5}{2}}-\rho^{\xi=n}_{012\frac{5}{2}}$, has the
largest value, followed by $n=0, n'=1$ mixings in p$_{3/2}$ and
p$_{1/2}$ ($\rho^{T=1}_{011\frac{3}{2}}$ and
$\rho^{T=1}_{011\frac{1}{2}}$ respectively). All of these are
normalized to the value of the largest element,
$\rho^{T=1}_{012\frac{5}{2}}$. The relative weight of each
off-diagonal contribution to the total isovector form factor can
change as the momentum transfer varies, and therefore it is necessary
in this case to compare the quantities
$f^{T=1}_{nn'lj}(q)\:\rho^{T=1}_{nn'lj}$. These are also shown in the
table for $q$ = 0.1, 0.5 and 1 fm$^{-1}$. Since the Coulomb monopole
operator between two s.h.o. wave functions $f_{nn'lj}(q)$ is
proportional to $q$ for off-diagonal contributions, they tend to
vanish as $q \to 0$. But for $q \neq 0$, although still relatively
small, off-diagonal contributions recover their prominent status
determined by the densities.

\subsection{Inter-comparisons of the full set of nuclei}

\begin{figure*}
\centering
\includegraphics[angle=-90,width=130mm]{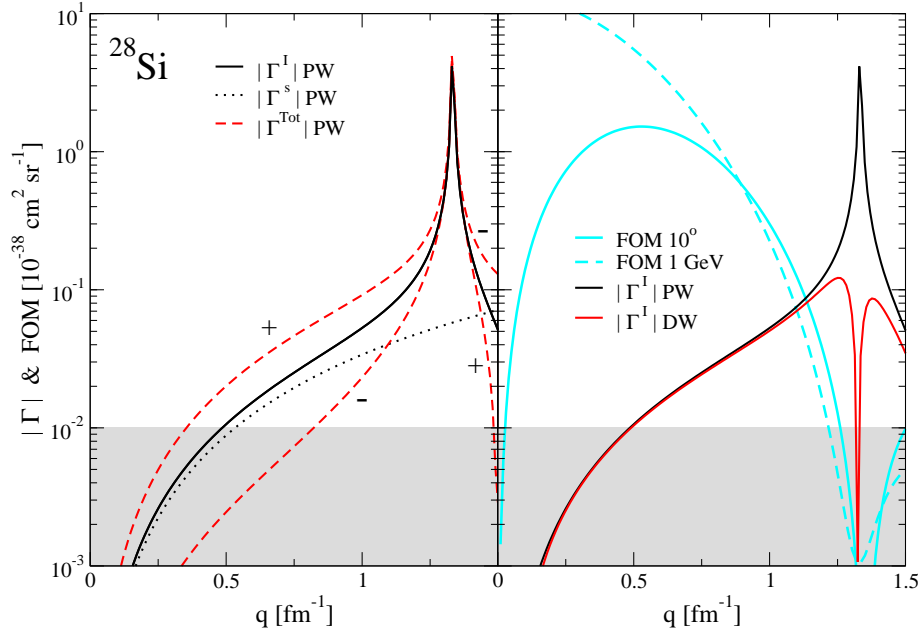}
\caption[]{ (Color online)
  Left-hand panel: total PV asymmetry deviation (dashed line) for
  $^{28}$Si from a HF(SLy4)+BCS($\Delta_{\pi,\nu}$= 1 MeV) calculation
  in PWBA with a strangeness contribution using two values of
  $\rho_s$, namely $\pm$1.5, together with the corresponding nucleon
  strangeness content (with $\rho_s$=1.5) contribution
  $\Gamma^{\text{s}}$ (dotted line). Total PV asymmetry for
  $\rho_s$=0, {\it i.e.,} the pure isospin-mixing contribution, is
  also shown (solid line). Right-hand panel: The latter curve again
  (solid dark line) and the same result but in DW with 1 GeV electrons
  (solid light line), together with the FOM corresponding to the total
  PV asymmetry for a fixed scattering angle of 10$^o$ (thick solid
  line) and for a fixed incident energy of 1 GeV (thick dashed
  line). The shaded region below $\Gamma$=0.01 is experimentally
  challenging, in contrast to the rest of the plot which, as discussed in the text,
  may be considered to be accessible with present technology.}
  \label{gamma_iso_strange_si28}
\end{figure*}

\begin{figure*}
\centering
\includegraphics[width=130mm]{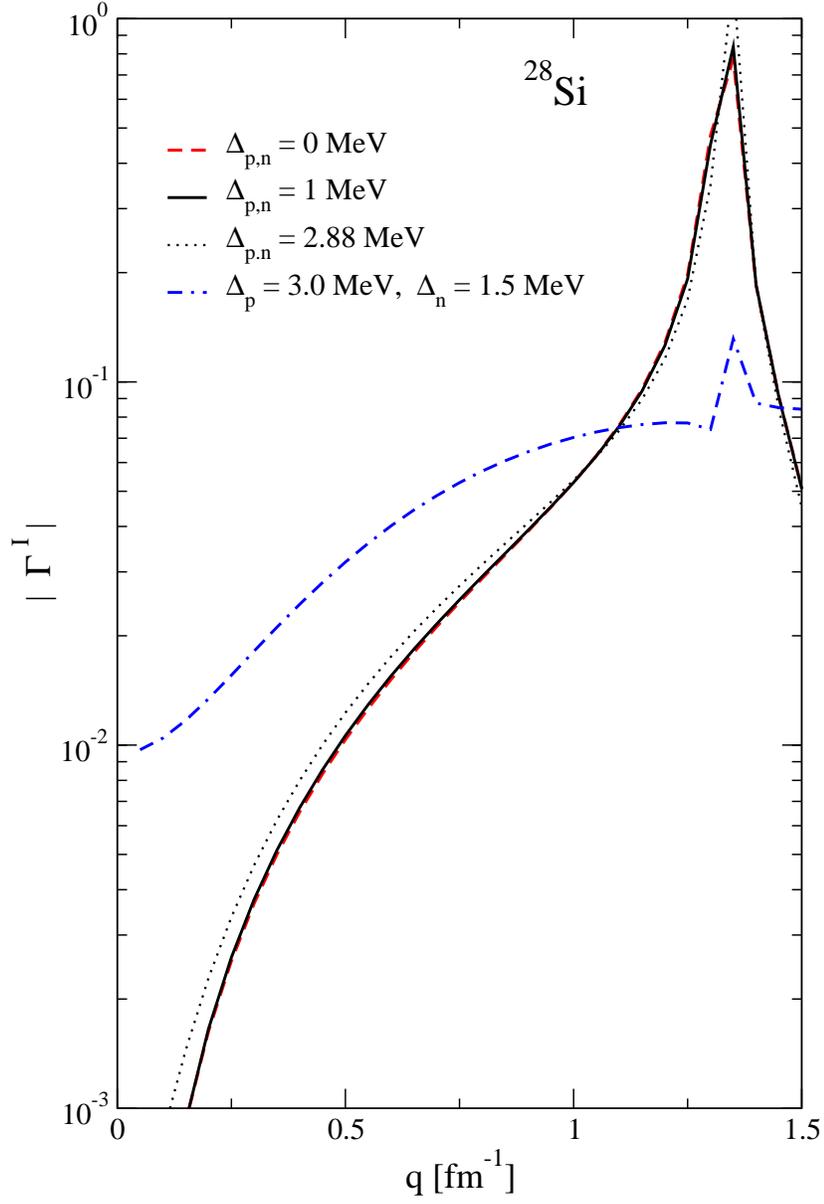}
\caption[]{(Color online) Isospin mixing contribution to the PV
  asymmetry deviation in $^{28}$Si from a HF(SLy4) (or HF+BCS)
  calculation for three values of the pairing gap: $\Delta_{\pi,\nu}$=
  0 MeV (solid line), $\Delta_{\pi,\nu}$= 1 MeV (dashed line) and
  $\Delta_{\pi,\nu}$= 2.88 MeV (dotted line). Also shown are results
  for different $\Delta_{\pi}$ and $\Delta_{\nu}$ values (dash-dotted
  line).}
\label{gamma_pairing_si28}
\end{figure*}

\begin{figure*}
\centering
\includegraphics[width=130mm]{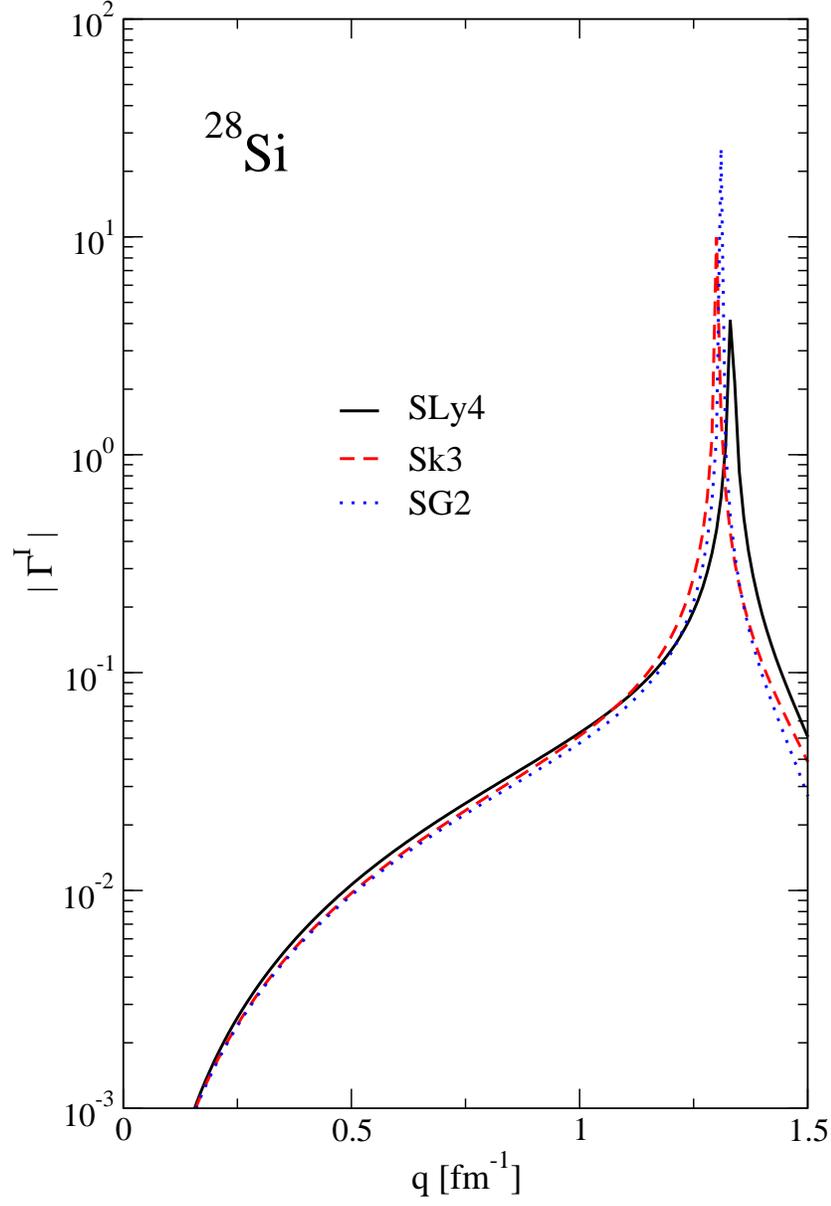}
\caption[]{(Color online) Isospin mixing contribution to the PV
  asymmetry deviation in $^{28}$Si from a HF+BCS calculation for three
  Skyrme forces: SLy4 (solid line), Sk3 (dashed line) and SG2 (dotted
  line).}
\label{gamma_forces_si28}
\end{figure*}

\begin{figure*}
\centering
\includegraphics[angle=-90,width=130mm]{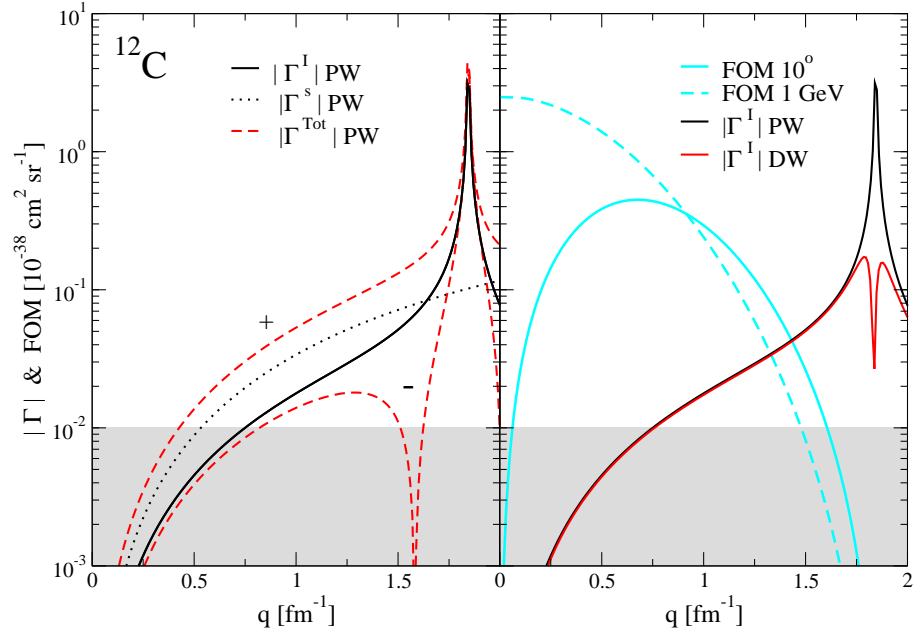}
\caption[]{(Color online) As for Fig.~\ref{gamma_iso_strange_si28}, but
  now for $^{12}$C.}
\label{gamma_iso_strange_c12}
\end{figure*}

\begin{figure*}
\centering
\includegraphics[angle=-90,width=130mm]{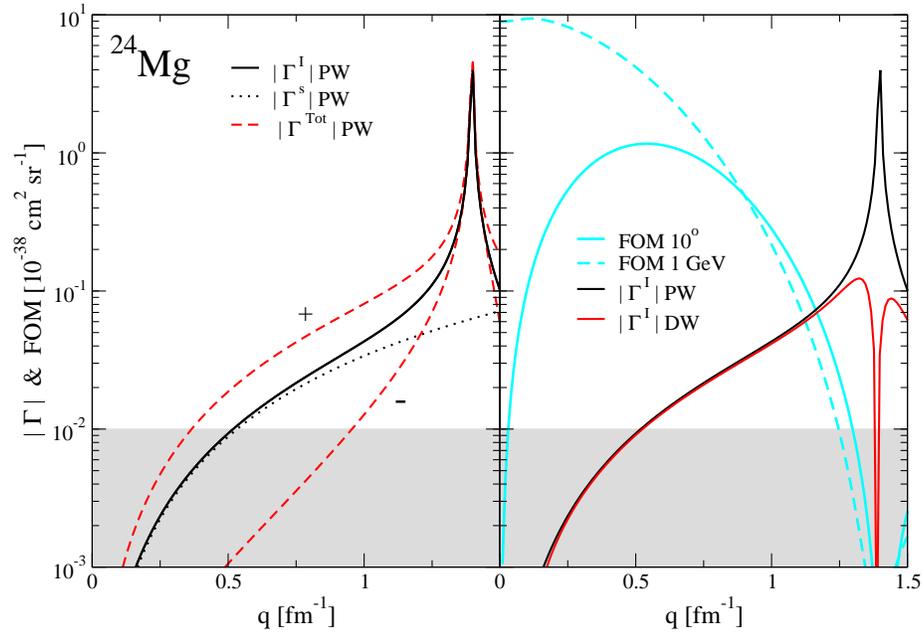}
\caption[]{(Color online) As for Fig.~\ref{gamma_iso_strange_si28}, but
  now for $^{24}$Mg.} \label{gamma_iso_strange_mg24}
\end{figure*}

\begin{figure*}
\centering
\includegraphics[angle=-90,width=130mm]{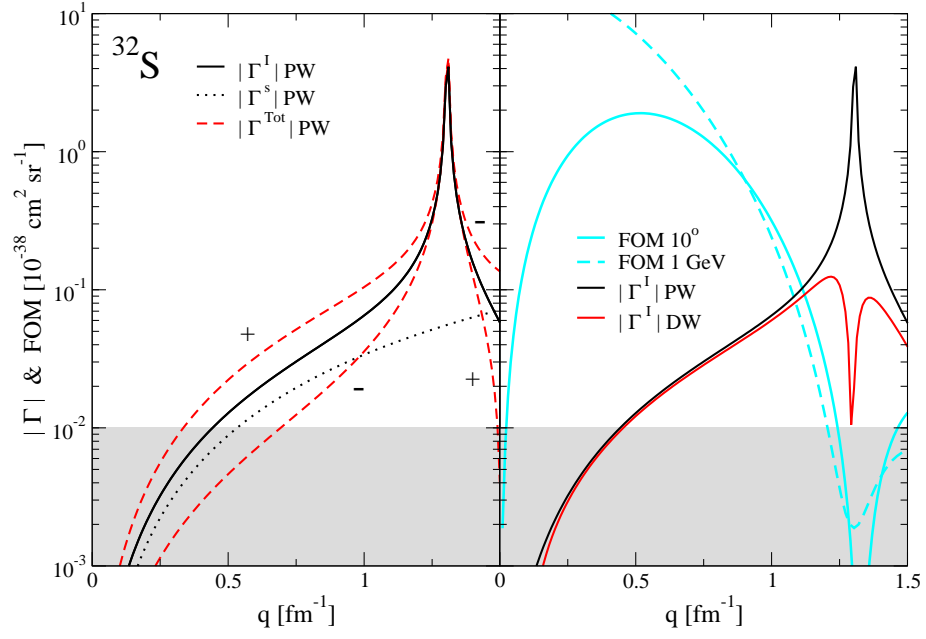}
\caption[]{(Color online) As for Fig.~\ref{gamma_iso_strange_si28}, but
  now for $^{32}$S.} \label{gamma_iso_strange_s32}
\end{figure*}

\begin{figure*}
\centering
\includegraphics[width=130mm]{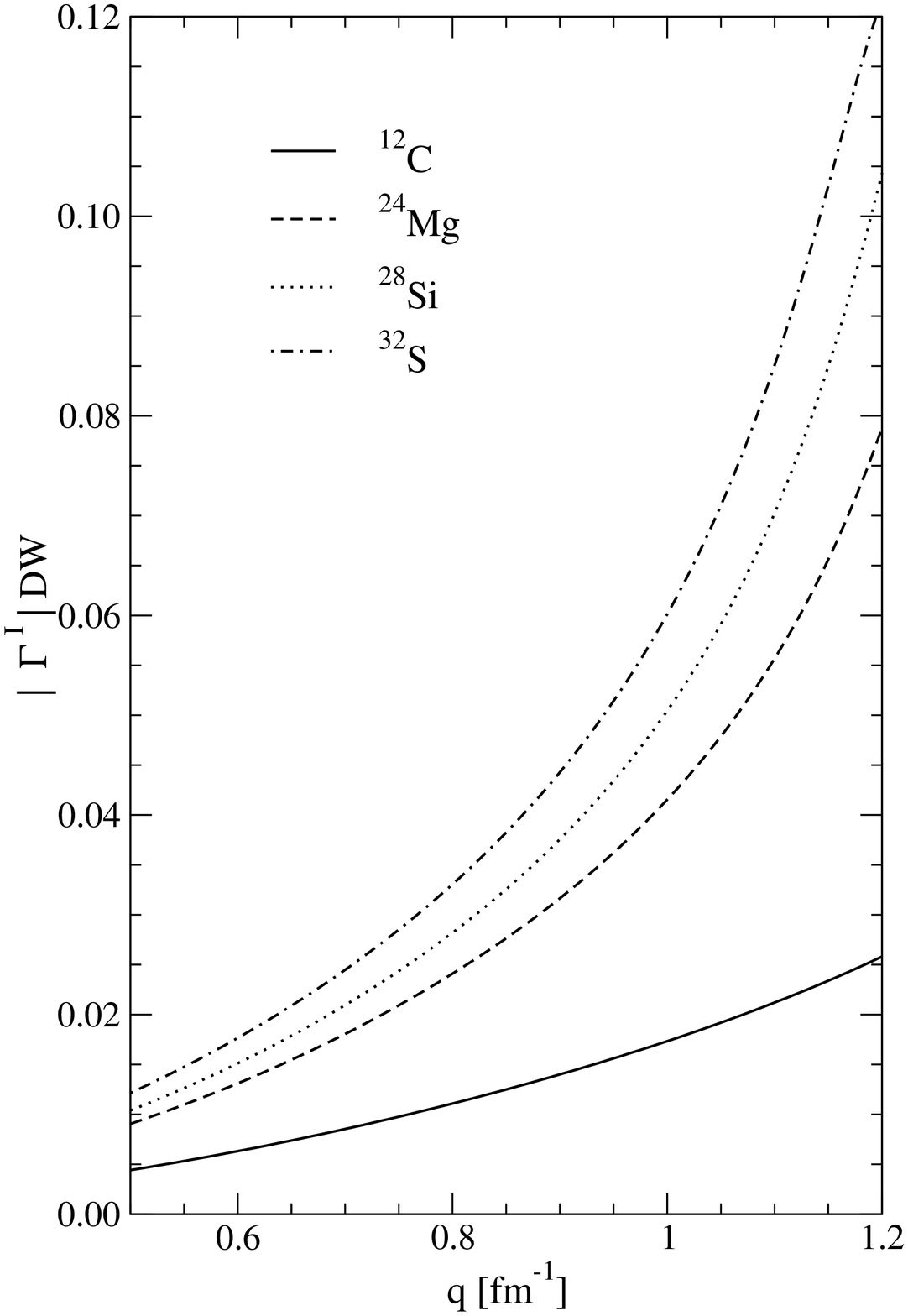}
\caption[]{Isospin mixing contribution to the PV asymmetry deviation
  $\Gamma^{\text{I}}$ for the four nuclei under study from a
  HF(SLy4)+BCS($\Delta_{\pi,\nu}$= 1 MeV) calculation in DW for 1 GeV
  electrons, in the momentum transfer region of most experimental
  interest. No strangeness contributions are included here.
}
\label{comp_gammaI_DW}
\end{figure*}

Now let us turn to general results at low momentum transfers shown in
Figs. \ref{gamma_iso_strange_si28} and
\ref{gamma_iso_strange_c12}--\ref{comp_gammaI_DW} for the full set of
four nuclei. In all cases the left-hand panels show results for the
total PV asymmetry deviation $|\Gamma|$ in PWBA for three values of
electric strangeness: results with no electric strangeness (solid
curve), with $\rho_s = +1.5$ (labeled $+$) and with $\rho_s = -1.5$
(labeled $-$). The strangeness contribution itself (with $|\rho_s| =
1.5$) is shown for reference as a dotted curve. In the right-hand
panels the $|\Gamma|$ obtained in plane-wave and distorted-wave cases
are compared (solid dark and light lines, respectively) and the FOM
obtained either at a fixed scattering angle of $\theta$ = 10$^{\circ}$
(in PWBA) or at fixed incident energy $\epsilon = 1$ GeV (in DW)
(solid and dashed lines, respectively) is presented. The former gives
a region around the peak where the FOM reaches reasonable values for
the PV deviation to be measured, whereas the latter just shows that
the smaller the scattering angle, the larger the FOM. When $q\to0$ the
FOM with fixed scattering angle behaves as $q^2$, whereas the FOM with
fixed incident energy behaves as a constant, this different behavior
for $q\to0$ being apparent in the figures.

To set the scale of ``doability'' we refer back to our earlier
discussions: the ability to measure the PV asymmetry is characterized
(at least) by the FOM and at constant angle this increases in going
from light nuclei such as $^4$He to a case such as $^{32}$S. Using
basic input concerning practical beam currents, electron
polarizations, detector solid angles and acceptable running times, in
past work (see DDS and also the discussions in Sec. 3.5.2 of
\cite{Mus94}), the lower limit was typically chosen to be 0.01, {\it
i.e.,} giving rise at least to a 1$\%$ change with respect to the
reference (Standard Model) value of the total PV asymmetry. In other
words, if something affects the PV asymmetry at (say) the few percent
level it can be regarded to be accessible with present
technology. Indeed, for elastic PV scattering from He 4\% has already
been reached \cite{Acha07} and upcoming experiments such as that on Pb
aim for even higher precision. Said another way, the peak value of the
FOM increases with nuclear species faster than $A$, while the
practical luminosity decreases roughly as $1/A$, making the peak FOM
$\times 1/A$ increase by roughly a factor of two in going from $^4$He
(see \cite{Mus94}) to $^{32}$S, and thus the ``doability'' actually
increases in going from light to heavier nuclei.  Accordingly, in the
present study in the figures we have shown where $|\Gamma|$ falls
below the 1\% level (namely, a bit more challenging than measurements
that have already been performed) to indicate the rough division
between what is presently within reach for PV electron scattering and
what will have to wait for future advances in technology;
specifically, in the figures we have indicated with shading where the
effects fall below 0.01 and are therefore likely to be inaccessible
for the foreseeable future. Table \ref{kinematicschoice} shows the
intervals of momentum transfer (in fm$^{-1}$) appropriate for the
measurement of the isospin-mixing part of the PV asymmetry deviations
(or, in other words, when $\rho_s=0$). These intervals have been
chosen so that the deviations are higher than 0.01 and at the same
time the FOM at fixed angle is no more than one order of magnitude
lower than the maximum value reached at the peak. They are
equivalently given in terms of incident electron energies (in MeV). In
the case of fixed incident energy, the intervals transform into
scattering angle ranges, also shown in the table (in degrees) for
$\epsilon = 1$ GeV. These angles are never smaller than 5$^{\circ}$,
which can be considered an experimental lower bound. As a rough
estimate, one can say that in the upper half of the intervals the
value of the FOM is very similar no matter what the combination of
incident energy and scattering angle used to get the correct transfer
momentum.

Note, furthermore, that the FOM is actually only the naive measure
({i.e.,} it characterizes $\delta {\cal A}/ {\cal A}$, as discussed
above), whereas, with the isospin-mixing effects increasing with $q$,
a somewhat large value of momentum transfer will actually be the
optimal one for experimental study. That is, a compromise must be
reached to choose a momentum transfer range where both the PV
asymmetry deviation and the FOM are as large as possible.  Given the
structure of the FOM at fixed angle, decreasing with the momentum
transfer due to its cross section dependence, the region of interest
corresponds to the first bump of the FOM graph, {\it i.e.,} between 0
and 1.5 fm$^{-1}$. Within this transfer momentum range, one has to
find a more restrictive region where the PV asymmetry deviation is
large enough to be measurable.  The present work is not intended to be
a detailed study of experimental possibilities and so we only use the
FOM for general guidance. In summary, everything above 10$^{-2}$ in
the figures can be considered to be measurable, while the shaded
region below this value will be experimentally more challenging.

In Figs.~\ref{gamma_iso_strange_si28} and
\ref{gamma_iso_strange_c12}--\ref{gamma_iso_strange_s32} we show in
the left-hand panel the PV asymmetry deviations for the three choices
of electric strangeness, $\rho_s = -1.5,$ 0 and +1.5, corresponding
roughly to the presently known range of acceptable values. Clearly for
positive versus negative electric strangeness the interplay between
strangeness and isospin mixing can be quite different: for negative
$G_E^{(s)}$ the two effects tend to cancel, whereas for positive
$G_E^{(s)}$ they interfere constructively. In the former case this
makes the predicted $|\Gamma|$ small, but still potentially
accessible, especially in the region at or slightly above $q=1$
fm$^{-1}$. In contrast, for positive electric strangeness the
asymmetry deviation typically rises above the 10\% value in regions
where the FOM is peaking --- clearly a significant modification of the
Standard Model PV asymmetry. One presumes that the $^4$He case is the
best suited to studies of electric strangeness, since isospin mixing
is predicted to be much smaller then; however, the interplay of
isospin mixing and strangeness seen here for the four nuclei under
investigation suggests that these cases could provide information not
only about the former effect, but also the latter.

In Figs.~\ref{gamma_iso_strange_si28} and
\ref{gamma_iso_strange_c12}--\ref{gamma_iso_strange_s32} we show in
the right-hand panel and in a reduced transfer momentum range the
isospin-mixing contribution to the asymmetry deviation computed within
DW from Eq.~(\ref{asymmetryratio_DW}) and once again in PWBA. The main
effect of the DW calculation is to smooth the divergence of the PWBA
asymmetry deviation by the appearance of a double-bumped
structure. Out of the region of the peaks, the differences between
PWBA and DW are seen to be negligible.

Finally, in order to emphasize how the isospin contribution to the PV
asymmetry changes as the atomic number increases, we show results from
a fully distorted calculation of it in Fig.~\ref{comp_gammaI_DW} for
the four nuclei under study. We restrict ourselves to the momentum
transfer region of most interest for future experiments, as suggested
in Table~\ref{kinematicschoice}, and the asymmetry deviation is shown
in a non-logarithmic scale. As can be seen in the figure, the isospin
contribution in this region of momentum transfer becomes higher as the
atomic number increases and typically $|\Gamma|$ goes from a few
percent for carbon to well over 10\% for the heavier cases. It should
be noted that this figure shows effects from isospin mixing alone: if
strangeness contributions of order those discussed above were to be
present then the total would be modified. For instance, if positive
electric strangeness contributions consistent with present
experimental limits were present in addition to the isospin-mixing
effects discussed in the present work, then the total asymmetry
deviations would be roughly twice those seen in the figure, namely,
comparatively large effects would be observed.

\section{Conclusions}

In the present work a new study has been undertaken of the effects
expected from isospin-mixing in nuclear ground-state wave functions on
elastic parity-violating electron scattering at momentum transfers
extending up to about 1.5 fm$^{-1}$. Four N=Z 0$^+$ nuclei have been
considered, $^{12}$C, $^{24}$Mg, $^{28}$Si, and $^{32}$S, each
expected to be very close to eigenstates of isospin with $T=0$ in
their ground states.  However, as first discussed in \cite{Don89}
(DDS), the Coulomb interaction occurs asymmetrically between pp and
pn/nn nucleon pairs in the nucleus, thereby giving rise to small
isospin mixing and thus the nuclear ground states considered here have
small admixtures with T$\neq$0.  While such effects are essentially
negligible in the parity-conserving cross section, they can play a
measurable role in the parity-violating asymmetry, and accordingly,
whether the focus is placed on isospin mixing itself or on how these
effects may confuse interpretations of the PV asymmetry in terms of
Standard Model tests or with respect to strangeness content in the
weak neutral current, it is important to evaluate their influence.

In the older work of DDS a limited-model-space shell model was
employed to estimate the isospin mixing and the mixing via the Coulomb
interaction was handled perturbatively via a simple two-level
approximation. Furthermore, in the study of DDS the idea of using PV
elastic electron scattering to determine ground-state neutron
distributions as in the case of $^{208}$Pb (which forms the basis of
the PREX experiment) was put forward; here the focus has been limited
to a few special N=Z nuclei and N$\neq$Z cases such as lead have not
been re-considered.

In the present work a self-consistent axially-symmetric mean-field
approximation with density-dependent effective two-body Skyrme
interactions, including Coulomb interactions between pp pairs, has
been used in direct determinations of the ground state wave
functions. The small differences between the proton and neutron
density distributions thereby obtained yield both isoscalar and
isovector ground-state Coulomb monopole matrix elements and produce
modifications in the PV asymmetry from the model-independent result
obtained in the absence of isospin-mixing and strangeness
contributions. Additionally, the effects of pairing in this mean-field
approximation have also been investigated, as have effects from
strangeness contributions in the single-nucleon form factors and from
subtle spin-orbit contributions in the Coulomb monopole operators.

A few of the important observations stemming from this study are
summarized below.

\begin{itemize}

\item In the present work one observes considerably larger effects
from isospin mixing than were found in DDS, especially since here
important matrix elements --- both diagonal and off-diagonal --- are
naturally included, whereas in the earlier study the restriction of
the shell-model space to a single major shell yielded a special
constraint on the $q$-dependences of the isovector matrix elements.

\item Specific influences of nuclear dynamics (different forces,
different pairing gaps) and of subtleties in the current operators
(spin-orbit effects) were investigated and seen not to affect the PV
asymmetry at low-$q$ significantly.

\item Results using either plane or distorted electron waves were
obtained and, for the relatively light nuclei considered in the
present work, their differences were seen to be small at low momentum
transfers except in the vicinity of a diffraction minimum where the
cross section is also small.

\item Kinematic ranges where potential future measurements might be
undertaken are discussed by studying both the deviations in the PV
asymmetry (the differences seen with/without isospin mixing) and the
experimental figure-of-merit. We have shown that the isospin mixing
effects considered in this work will have a measurable effect on the
asymmetries, even for the light nuclei considered.

\item Furthermore, in going from the lightest N=Z nuclei to heavier
cases one sees that the asymmetry deviations increase making the
isospin-mixing effects all the more evident.

\item In exploring the interplay between the isospin-mixing effects
and those effects that may arise from electric strangeness
contributions one sees an interesting constructive/destructive
interference scenario: with positive electric strangeness the two
contributions add at low momentum transfers, whereas when negative
they subtract. Presently the sign of the electric strangeness form
factor is unknown (the form factor is, in fact, consistent with zero)
and so such interferences may provide information on isospin mixing,
namely the main focus of the present work, but also on strangeness.

\end{itemize}

\begin{acknowledgments}
  This work was supported by Ministerio de Ciencia e Innovaci\'on
  (Spain) under Contracts No. FIS2005-00640 and No. FIS2008-01301.
  O.M. thanks Ministerio de Ciencia e Innovaci\'on (Spain) for
  financial support. J.M.U. acknowledges support from INTAS Open Call
  grant No 05-1000008-8272, and Ministerio de Ciencia e Innovaci\'on
  (Spain) under grants FPA-2007-62616 and FPA-2006-07393, and UCM and
  Comunidad de Madrid under grant Grupo de F\'isica Nuclear (910059).
  This work was also supported in part (TWD) by the U.S. Department of
  Energy under contract No. DE-FG02-94ER40818. TWD also wishes to
  thank UCM-GRUPOSANTANDER for financial support at the Universidad
  Complutense de Madrid.

\end{acknowledgments}

\newpage

\begin{table}[t]
  \caption{ Theoretical (deformed HF or HF+BCS) and experimental \cite{Rag89} 
intrinsic proton quadrupole moments $Q_{0\:p}$ [fm$^2$] of the ground-state 
rotational band in each of the isotopes under study, together with the 
corresponding pairing gap $\Delta_{\pi,\nu}$
    used in the HF+BCS (coul.+pair.) calculation.}
\label{tabledef}
\begin{tabular}{ccccc}\cr
 Isotope  &  $\quad \Delta_{\pi,\nu}$ = 0 \quad &  $\quad \Delta_{\pi,\nu}$ 
= 1 MeV \quad &  $\quad \Delta_{\pi,\nu}$ from mass diff. \quad &\quad Exp. 
\cite{Rag89}
\cr
\hline
\cr
$^{12}$C  & $\sim$0 & $\sim$0  & $\sim$0  & -  \cr
$^{24}$Mg & 56.67 & 54.70  & 38.54 & 58.1 \cr
$^{28}$Si & -45.35 & -43.41  & -28.62 & -57.75  \cr
$^{32}$S  & $\sim$0 & $\sim$0  & $\sim$0  & -  \cr
\cr
\end{tabular}
\end{table}

\begin{table}[t]
  \caption{ Main single-particle ($l,j$) components of the total
    HF+BCS ground-state isovector density
    $\rho^{T=1}_{nn'lj}=\rho^{\xi=p}_{nn'lj}-\rho^{\xi=n}_{nn'lj}$ and
    Coulomb monopole form factor $f_{nn'lj}(q)\:\rho^{T=1}_{nn'lj}$
    (at $q=$ 0.1, 0.5 and 1 fm$^{-1}$) of $^{28}$Si, for different
    combinations of radial quantum numbers $n,n'$. Results are
    normalized to the largest contribution and only the absolute
    values are given, not the signs. The largest value for each
    ($l,j$) component is written in boldface. }
\label{multipolar}
\begin{tabular}{cccccccc}\cr
 &  &  &  & $\quad \rho^{T=1}_{nn'lj} \quad$  & \multicolumn{3}{c}{\quad 
$f_{nn'lj}\:\rho^{T=1}_{nn'lj}$}  \cr
\cline{6-8} \cr
 $\quad l \quad$  & $\quad j \quad$ & $\quad  n \quad$  & $\quad  n' \quad$
 &   & $\quad q$ = 0.1 fm$^{-1} \quad$ & $\quad q$ = 0.5 fm$^{-1} \quad$
 & $\quad q$ = 1 fm$^{-1} \quad$ \cr
\cr
\hline
\cr
   &     & 0 & 0 & 0.058 & \textbf{0.208} & 0.229 & 0.145 \cr
   &     & 0 & 1 & 0.168 & 0.005 & 0.136 & \textbf{0.346} \cr
0  & 1/2 & 0 & 2 & 0.052 & 0.000 & 0.005 & 0.048 \cr
   &     & 1 & 1 & 0.028 & 0.100 & 0.079 & 0.024 \cr
   &     & 1 & 2 & \textbf{0.203} & 0.011 & \textbf{0.244} & 0.304 \cr
\cr
  &      & 0 & 0 & 0.093 &  \textbf{0.335} & 0.309 & 0.078 \cr
  &      & 0 & 1 & \textbf{0.349} & 0.013 & \textbf{0.329} &  \textbf{0.555} \cr
1  & 1/2 & 0 & 2 & 0.048 & 0.000  & 0.006 & 0.048 \cr
  &      & 1 & 1 & 0.093 & 0.330  & 0.217 & 0.047 \cr
  &      & 1 & 2 & 0.018 & 0.001  & 0.023 & 0.018 \cr
\cr
  &      & 0 & 0 & 0.184 & \textbf{0.662} & \textbf{0.610} & 0.016 \cr
  &      & 0 & 1 & \textbf{0.628} & 0.024 & 0.592 & \textbf{1.000}  \cr
1  & 3/2 & 0 & 2 & 0.137 & 0.000 & 0.018 & 0.137  \cr
  &      & 1 & 1 & 0.186 & 0.659 &0.434 & 0.094  \cr
  &      & 1 & 2 & 0.031 & 0.002 & 0.040 & 0.032  \cr
\cr
  &      & 0 & 0 & 0.044 & \textbf{0.156} & \textbf{0.119} & 0.007  \cr
  &      & 0 & 1 & \textbf{0.061} & 0.003 & 0.061 & \textbf{0.060}   \cr
2  & 3/2 & 0 & 2 & 0.008 & 0.000 & 0.001 & 0.007  \cr
  &      & 1 & 1 & 0.001 & 0.004 & 0.002 & 0.000  \cr
  &      & 1 & 2 & 0.002 & 0.000 & 0.002 & 0.001  \cr
\cr
  &      & 0 & 0 & 0.142 & 0.506 & 0.386 & 0.024  \cr
  &      & 0 & 1 & \textbf{1.000} & 0.045 & \textbf{1.000} & \textbf{0.986}  \cr
2  & 5/2 & 0 & 2 & 0.048 & 0.000 & 0.008 & 0.044  \cr
  &      & 1 & 1 & 0.284 & \textbf{1.000} & 0.541 & 0.075  \cr
  &      & 1 & 2 & 0.379 & 0.003 & 0.049 & 0.021  \cr
\cr
\end{tabular}
\end{table}

\begin{table}[t]
  \caption{ Momentum transfer intervals
    (in fm$^{-1}$) appropriate for the measurement of the
    isospin-mixing contribution ({\it i.e.,} $\rho_s=0$) to the PV
    asymmetry deviation, chosen so that it is higher than 0.01 and the
    FOM at $\theta$=10$^\circ$ is no more than one order of magnitude
    lower than its maximum value (see
    Figs.~\ref{gamma_iso_strange_si28}, \ref{gamma_iso_strange_c12} --
    \ref{gamma_iso_strange_s32}).  Corresponding ranges of incident
    energy $\epsilon$ at $\theta$=10$^{\circ}$ (in MeV) and scattering
    angle $\theta$ at E=1 GeV (in degrees) are also given.}
\label{kinematicschoice}
\begin{tabular}{cc}\cr
Isotope  & \quad for $\Gamma^{\text{I}}$ \quad \cr
\hline
\cr
$^{12}$C  & 0.74 $\le$ q $\le$ 1.42   \cr
          & 838 $\le$ $\epsilon$ $\le$ 1607  \cr
          & 8.38 $\le$ $\theta$ $\le$ 16.07  \cr
\cr
$^{24}$Mg & 0.51 $\le$ q $\le$ 1.10   \cr
          & 577 $\le$ $\epsilon$ $\le$ 1245  \cr
          & 5.77 $\le$ $\theta$ $\le$ 12.45   \cr
\cr
$^{28}$Si & 0.48 $\le$ q $\le$ 1.08   \cr
          & 543 $\le$ $\epsilon$ $\le$ 1222   \cr
          & 5.43 $\le$ $\theta$ $\le$ 12.22   \cr
\cr
$^{32}$S  & 0.45 $\le$ q $\le$ 1.05   \cr
          & 509 $\le$ $\epsilon$ $\le$ 1188   \cr
          & 5.09 $\le$ $\theta$ $\le$ 11.88   \cr
\cr
\end{tabular}
\end{table}

\newpage

\begin{thebibliography}{00}

\bibitem{Fei75} G. Feinberg, Phys. Rev. D 12 (1975) 3575.

\bibitem{Wal77} J. D. Walecka, Nucl. Phys. A 258 (1977) 349.

\bibitem{Don79} T. W. Donnelly and R. D. Peccei, Phys. Rep. 50 (1979) 1.

\bibitem{Mus94} M. J. Musolf, T. W. Donnelly, J. Dubach, S. J. Pollock,
S. Kowalski and E. J. Beise,  Phys. Rep. 239 (1994) 1.

\bibitem{Don89} T. W. Donnelly, J. Dubach and I. Sick, Nucl. Phys. A 503
(1989) 589.

\bibitem{Hor01} C. J. Horowitz, S. J. Pollock, P.A. Souder and R. Michaels,
Phys. Rev. C 63 (2001) 025501.

\bibitem{Aue83} N. Auerbach, Phys. Rep. 98 (1983) 273.

\bibitem{Acha07} A. Acha, et al., Phys. Rev. Lett. 98 (2007) 032301.

\bibitem{PREX} http://hallaweb.jlab.org/parity/prex/

\bibitem{Lhu08} D. Lhuillier  Prog. Nucl. Part. Phys. 61 (2008) 183.

\bibitem{Ama96} J. E. Amaro, J. A. Caballero, T. W. Donnelly, A. M. Lallena, 
E. Moya de Guerra and Ud\'{\i}as, Nucl. Phys. A 602 (1996) 263.

\bibitem{Ama96_2} J. E. Amaro, J. A. Caballero, T. W. Donnelly and 
E.  Moya de Guerra, Nucl. Phys. A 611 (1996) 163.

\bibitem{Jes98} S. Jeschonnek and T. W. Donnelly, Phys. Rev. C 57 (1998) 2438.

\bibitem{Mus94a} M. J. Musolf, R. Schiavilla and T. W. Donnelly, Phys. Rev. C 50 (1994) 2173. 

\bibitem{Hoh76} G. H\"ohler, et al., Nucl. Phys. B 114 (1976) 505.

\bibitem{sly4} E. Chabanat, P. Bonche, P. Haensel, J. Meyer and R. Schaeffer,
Nucl. Phys. A 635 (1998) 231.

\bibitem{vautherin} D. Vautherin and D. M. Brink,  Phys. Rev. C 5 (19272) 626;
D. Vautherin, Phys. Rev. C 7 (1973) 296.

\bibitem{Moy91} E. Moya de Guerra, P. Sarriguren, J. A. Caballero, M.
  Casas and D. W. L. Sprung, Nucl. Phys. A 529 (1991) 68.

\bibitem{Moy86} E. Moya de Guerra, Phys. Rep. 138 (1986) 293.

\bibitem{Zar77} A. Zaringhalam and J. W. Negele, Nucl. Phys. A 288 (1977)  417.

\bibitem{Alv05} R. \'Alvarez-Rodr\'{\i}guez, E. Moya de Guerra and
P. Sarriguren, Phys. Rev. C 71 (2005) 044308.

\bibitem{Hor98} C. J. Horowitz, Phys. Rev. C 57 (1998) 3430.

\bibitem{Ruf82} G. Rufa, Nucl. Phys. A 384 (1982) 273.

\bibitem{Ant05} A. N. Antonov, D.N. Kadrev, M.K. Gaidarov, E. Moya de
  Guerra, P. Sarriguren, J.M. Udias, V.K. Lukyanov, E.V. Zemlyanaya and
  G.Z. Krumova, Phys. Rev. C 72 (2005) 044307.

\bibitem{Sick70}
I.~Sick and J. S. McCarthy, Nucl. Phys. A 150 (1970) 631.

\bibitem{Jansen72}
J. A. Jansen, R. T. Peerdeman and C.~deVries, Nucl. Phys. A 188 (1972) 337.

\bibitem{Fey73}
G. Fey, Thesis, TH Darmstadt, 1973.

\bibitem{Cardman80}
L. S. Cardman, J. W. Lightbody, S.~Penner, W. P. Trower and S. F.
Williamson, Phys. Lett. B 91 (1980) 203.

\bibitem{Reuter81}
W.~Reuter, Thesis KPH Mainz, 1981.

\bibitem{Li74}
G. C. Li, I.~Sick and M. R. Yearian, Phys. Rev. C 9 (1974) 1861.

\bibitem{Lees76}
E. W. Lees, C. S. Curran, T. E. Drake, W. A. Gillespie, A. Johnson
and R. P. Singhal, J. Phys. G 2 (1976) 105.

\bibitem{Li71a}
G. C. Li, I.~Sick and M. R. Yearian, Phys. Lett. B 37 (1971) 282.

\bibitem{Aud03} G. Audi, O. Bersillon, J. Blachot and A. H. Wapstra,
Nucl. Phys. A 729 (2003) 3.

\bibitem{Rag89} P. Raghavan, Atomic and Nuclear Data Tables 42 (1989) 189; 
N.J. Stone, Table of Nuclear Moments (2001)  www.nndc.bnl.gov/nndc/stone$\_$moments

\bibitem{Fir99} Table of Isotopes, 8 th. ed., 1999 update (eds. R.B.
  Firestone and V.S. Shirley) (Wiley Interscience, 1999)

\bibitem{Fyn05} H. Fynbo, et al., Science 433 (2005) 136.

\bibitem{Leinweber06} D. B. Leinweber, S. Boinepalli, A. W. Thomas, P. Wang,
A. G. Williams, R. D. Young, J. M. Zanotti and J. B. Zhang, Phys.
Rev. Lett. 97 (2006) 022001.

\bibitem{Kubis06} B. Kubis and R. Lewis, Phys. Rev. C 74 (2006) 015204.

\bibitem{Rama94} S. Ramavataram, E. Hadjimichael and T. W. Donnelly,
Phys. Rev. C 50 (1994) 1175.

\bibitem{Viviani07} M. Viviani, R. Schiavilla, B. Kubis, R. Lewis, L.
  Girlanda, A. Kievsky, L.E. Marcucci and S. Rosati, Phys. Rev.  Lett.
  99 (2007) 112002.

\bibitem{Zhou07} H. Q. Zhou, C. W. Kao and S. N. Yang, Phys. Rev. Lett. 99
(2007) 262001.

\bibitem{sk3} M. Beiner, H. Flocard, N. Van Giai and P. Quentin,
Nucl. Phys. A 238 (1975) 29.

\bibitem{sg2} N. Van Giai and H. Sagawa, Phys. Lett. B 106 (1981) 379.

\end{thebibliography}
\end{document}